\documentclass[a4paper,fleqn]{cas-dc}

\usepackage[square,numbers,sort&compress]{natbib}
\usepackage{csquotes}
\usepackage{xcolor}
\usepackage[colorinlistoftodos,prependcaption]{todonotes}
\usepackage[inline]{enumitem}
\usepackage{xspace}
\usepackage{subcaption}
\usepackage{pifont}

\usepackage{tablefootnote}
\usepackage{url}
\usepackage{float}
\usepackage{balance}
\usepackage{longtable}
\usepackage[euler]{textgreek}
\usepackage{acronym} 
\usepackage{threeparttable} 
\usepackage{multicol}
\usepackage{amssymb}%
\usepackage{pifont}% http://ctan.org/pkg/pifont
\newcommand{\cmark}{\ding{51}}%
\newcommand{\xmark}{\text{\ding{55}}}

\def\tsc#1{\csdef{#1}{\textsc{\lowercase{#1}}\xspace}}
\tsc{WGM}
\tsc{QE}
\tsc{EP}
\tsc{PMS}
\tsc{BEC}
\tsc{DE}
\begin{document}
\let\WriteBookmarks\relax
\def\floatpagepagefraction{1}
\def\textpagefraction{.001}

\makeatletter
\renewcommand{\@oddfoot}{\hfill \textit{Accepted in Springer} \hfill}
\renewcommand{\@evenfoot}{\@oddfoot}
\makeatother

% Short title
\shorttitle{Image and Point-cloud Classification for Jet Analysis in High-Energy Physics: A survey}

% Short author
\shortauthors{H. Kheddar et~al.}

\title [mode = title]{Image and Point-cloud Classification for Jet Analysis in High-Energy Physics: A survey}                      

\vskip2mm

\author[1]{Hamza Kheddar\corref{cor1}}
[orcid=0000-0002-9532-2453]
\ead{kheddar.hamza@univ-medea.dz}
\credit{Conceptualization; Methodology; Data Curation; Resources; Investigation; Visualization;  Writing original draft; Writing, review, and editing}

\author[2]{Yassine Himeur}
[orcid=0000-0001-8904-5587]
\ead{yhimeur@ud.ac.ae}
\credit{Methodology; Resources; Visualization; Investigation; Writing – review and editing, Supervision}

\author[3,4]{Abbes Amira}
[orcid=0000-0003-1652-0492]
\ead{aamira@sharjah.ac.ae}
\credit{Methodology; Resources; Visualization; Investigation; Writing – review and editing, Supervision}

\author[5,6]{Rachik Soualah}
[orcid=0000-0003-0124-3410]
\ead{rachik.soualah@ku.ac.ae/rsoualah@cern.ch}
\credit{Methodology; Resources; Visualization; Investigation; Writing – review and editing, Supervision}

\address[1]{LSEA Laboratory, Electrical Engineering Department, University of Medea, Algeria}

\address[2]{College of Engineering and Information Technology, University of Dubai, Dubai, UAE}

\address[3]{Department of Computer Science, University of Sharjah, UAE}
\address[4]{Institute of Artificial Intelligence, De Montfort University, Leicester, United Kingdom}

\address[5]{Department of Physics, Khalifa University of Science and Technology, P.O. Box 127788, 127788, Abu Dhabi, United Arab Emirates}

\address[6]{The International Center for Theoretical Physics (ICTP), Strada Costiera, 11, I-34151 Trieste, Italy}

\tnotetext[1]{The first is the corresponding author.}

% Here goes the abstract
\begin{abstract}
Nowadays, there has been a growing trend in the field of high-energy physics (HEP), in both its experimental and phenomenological studies, to incorporate machine learning (ML) and its specialized branch, deep learning (DL). This review paper provides a thorough illustration of these applications using different ML and DL approaches. The first part of the paper examines the basics of various particle physics types and establishes guidelines for assessing particle physics alongside the available learning models. Next, a detailed classification is provided for representing Jets that are reconstructed in high-energy collisions, mainly in proton-proton collisions at well-defined beam energies. This section covers various datasets, preprocessing techniques, and feature extraction and selection methods. The presented techniques can be applied to future hadron-hadron colliders (HHC), such as the high-luminosity LHC (HL-LHC) and the future circular collider - hadron-hadron (FCC-hh). The authors then explore several AI techniques  analyses designed specifically for both image and point-cloud (PC) data in HEP. Additionally, a closer look is taken at the classification associated with Jet tagging in hadron collisions. In this review, various state-of-the-art (SOTA) techniques in ML and DL are examined, with a focus on their implications for HEP demands. More precisely, this discussion addresses various applications in extensive detail, such as Jet tagging, Jet tracking, particle classification, and more. The review concludes with an analysis of the current state of HEP using DL methodologies. It highlights the challenges and potential areas for future research, which are illustrated for each application.

\end{abstract}

% Use if graphical abstract is present
% \begin{graphicalabstract}
% \includegraphics{figs/grabs.pdf}
% \end{graphicalabstract}

% Research highlights
%\begin{highlights}
%\item Research highlights item 1
%\item Research highlights item 2
%\item Research highlights item 3
%\end{highlights}

% Keywords
% Each keyword is seperated by \sep

\maketitle

\begin{table*}[]
\centering
{\small \section*{Abbreviations list}}
\begin{multicols}{3}
\footnotesize
\begin{acronym}[AWGN] 
\acro{AE}{auto-encoder}
\acro{AI}{artificial intelligence}
%\acro{AMS}{approximate median significance}
\acro{ANN}{artificial neural network}
\acro{AUC}{area under curve}
\acro{BDT}{boosted decision trees}
\acro{BIP}{boost-invariant polynomial}
\acro{BRNN}{bidirectional RNN}
\acro{CERN}{conseil Européen pour la recherche nucléaire}
\acro{CG}{Clebsch-Gordan}
\acro{CGENN}{Clifford group equivariant neural networks}
\acro{CMS}{compact muon solenoid}
\acro{CMSSM}{constrained minimal super-symmetric standard model}
\acro{CNN}{convolutional neural network}
\acro{DL}{deep learning}
\acro{DLA}{dataset link availability}
\acro{DNN}{deep neural network}
\acro{EG}{event generator}
\acro{EGNN}{equivariant graph neural networks}
\acro{FAR}{false alarm rate}
\acro{FGSM}{fast gradient sign method}
\acro{FL}{federated learning}
\acro{FPR}{false-positive rate}
\acro{FST}{Fubini-study tensor}
\acro{GAN}{generative adversarial network}
\acro{GMM}{Gaussian mixture model }
\acro{GNN}{graph neural network}
\acro{GPU}{graphics processing unit}
\acro{GRU}{gated recurrent unit}
\acro{HCAL}{hadronic calorimete}
\acro{HEP}{high-energy physics}
\acro{HPC}{high-performance computing}
\acro{KNN}{K-nearest neighbors}
\acro{L-GATr}{Lorentz geometric algebra Transformer}
\acro{LGEB}{Lorentz group equivariant block}
\acro{LGN}{Lorentz group network}
\acro{LHC}{large hadron collider}
\acro{LLM}{large language model}
\acro{LSTM}{long short-term memory}
\acro{MIParT}{more-interaction particle Transformer}
\acro{ML}{machine learning}
\acro{MLP}{multi-Layer perceptron}
\acro{MSE}{mean squared error}
\acro{PC}{point cloud}
\acro{PCA}{ principal component analysis}
\acro{PCT}{point cloud Transformer}
\acro{$p_T$}{transverse momentum}
\acro{PFN}{particle flow network}
\acro{PFN-ID}{particle flow network with ID}
\acro{PLP}{primary lund plane}
\acro{P-MHA}{pairwise multi-head attention}
\acro{QCD}{quantum chromodynamics}
\acro{QGP}{quark-gluon plasma}
\acro{QML}{quantum machine learning}
\acro{RecNN}{recursive neural network}
\acro{ReLU}{rectified linear unit}
\acro{RL}{reinforcement learning}
\acro{RNN}{recurrent neural network}
\acro{ROC}{receiver operating characteristic}
\acro{SHAP}{shapley additive explanations}
\acro{SM}{standard model}
\acro{SOTA}{state-of-the art}
\acro{SVM}{support vector machine}
\acro{TL}{transfer learning}
\acro{TPR}{true positive rate}
\acro{t-SNE}{t-distributed stochastic neighbor embedding}
\acro{UMAP}{uniform manifold approximation and projection}
\acro{ViT}{vision Transformers}
\end{acronym}

\end{multicols}
\end{table*}

\section{Introduction}
\Ac{HEP} is an attracting and delicate branch of physics that manifests at the microscopic scale and which explores the fundamental building blocks of the universe and forces that govern their interactions at incredibly high energies under extremely intense conditions \cite{zhou2023exploring,belis2024machine}. In this field many sophisticated instruments and tools with large particle accelerators, like the current CERN-LHC (located near the French and Swiss border), to study matter at energy levels that are otherwise unattainable to reach with conventional methods.  These gigantic machines accelerate subatomic particles at nearly the speed of light and then smash them together, creating energy densities analogous of the early moments after the Big Bang \cite{guo2024mlanalysis,tani2024comparison}. By studying the collisions generated in these accelerators setups, it could be possible to track and evaluate rare particles that have a very short life time. This important study with the accumulated big data at higher collider luminosity values offers an improved understanding of the basic anatomy of different physics process and their topologies \cite{durante2019applied,kansal2023evaluating}.

The \ac{SM} is the present theoretical framework that describes the elementary particles and their interactions \cite{bauer2023quantum}. Despite its tremendous success  in explaining many phenomena in nature, several mysteries remain unsolved, such as matter antimatter asymmetry, the nature of dark matter (DM), the neutrino mass and  the hierarchy problem and many other open questions \cite{bramante2024dark}.
Furthermore, it is worth noting that besides its deep investigation about the Universe puzzles, \ac{HEP} has demonstrated significant practical utility when used with advanced technologies \cite{he2023high}. As a matter of fact, the development of many techniques and technologies in this sector has driven notable progress in medical imaging \cite{kashkooli2023ultrasound,sohail2023xai}, radiation therapy \cite{graeff2023emerging,kraan2024technological}, and materials research \cite{dorigo2023toward,bilici2023monte}.

The data acquisition system of \ac{LHC} stores the data on tape using grid computing facilities, it can be disseminated for offline analysis aimed at extracting information concerning particle trajectories formed within the detectors. These trajectories contain concealed details about numerous particle characteristics. {Jets are reconstructed by combining information from multiple detector subsystems, primarily calorimeters and trackers. The calorimeters (electromagnetic and hadronic) play a central role by capturing the energy deposits from both neutral and charged particles. These deposits are clustered using algorithms such as anti-\( k_t \), which group the energy into Jets based on angular proximity in \((\eta, \phi)\)-space. While tracking systems provide detailed momentum and charge information for individual charged particles, they cannot detect neutral particles, such as photons or neutrons. Therefore, the calorimeter serves as the primary tool for measuring the total energy of the Jet. This reconstruction process ensures that Jets are defined as comprehensive objects representing the full range of particle constituents, crucial for subsequent analyses in HEP experiments \cite{di2023reconstructing}.}

Computer vision techniques become relevant and play a crucial role during the analysis of offline data. Specifically, in the realm of \ac{HEP} data analysis, \ac{ML} algorithms have found success, leading to significant enhancements in event classification performance when contrasted with traditional methods rooted in expert understanding. Techniques like \ac{BDT}, shallow neural networks, and similar approaches have been employed in \ac{HEP} data analysis. More recently, \ac{DNN} or \ac{DL} have gained widespread adoption due to their applicability to intricate data structures such as images, videos, natural language, or sensor data. There are ongoing investigations into applying \acp{DNN} for analyzing granular details like particle positions and momentum as they traverse the detector. This has shown increased effectiveness in selecting signal events compared to \ac{ML} algorithms employing conventional feature variables rooted in physics knowledge \cite{kim2023multi}.

\subsection{Motivation}

In \ac{HEP}, a \textit{track} typically refers to the trajectory or path followed by a charged particle as it moves through a particle detector. \ac{HEP} experiments often involve the collision of high-energy particles, such as those produced in particle accelerators like \ac{LHC}. When these particles collide, they produce various other particles as a result of the collisions. These newly created particles then pass through several sub-detectors where each designed to measure their corresponding properties. Each charged particle leaves behind a trace or \textit{track } as it interacts with the detector's various components, such as tracking chambers or silicon detectors. These tracks provide information about the particle's momentum, charge, and the path it took through the detector. Analyzing these tracks is very crucial for understanding the physics of the collisions and for identifying the types of particles produced.

The reconstruction of particle tracks involves sophisticated algorithms and software that piece together the recorded data from various detector components to reconstruct the paths of the particles accurately. Then, the reconstructed tracks are essential for a wide range of analyses in \ac{HEP}, including the discovery of new particles, the measurement of particle properties, and the investigation of fundamental forces and interactions in the universe.

However, in \ac{HEP} experiments, there are always chances for high background
contributions or events that are not of primary interest and can eventually mimic the physics signal and moreover can interfere along the physics collision. The background sources could be the electronic components in the different detector systems, when high-energetic particles pass through the material budget of the detector, they can also generate secondary tracks through different interactions, and possible decay modes.

In the light of the aforementioned phenomena and challenges, treating tracks / Jets in \ac{HEP} as image  {or \ac{PC}}-like data for processing and analysis is a useful approach, especially when dealing with the output from particle detectors. Hence, \ac{ML} and \ac{DL} play vital roles in \ac{HEP} experiments. They serve the following purposes: i) Identifying and classifying particles by analyzing their tracks and energy deposits in detectors, thereby enhancing precision and identification speed, ii) assisting in the accurate reconstruction of particle tracks from detector data, particularly in complex environments with numerous particles and interactions, iii) enabling efficient data analysis schemes, one can sift through extensive datasets to pinpoint rare or noteworthy events or particles, iv) detecting anomalies or unexpected patterns in the recorded data, which could potentially signify the existence of new particles and physics beyond the \ac{SM}, among other applications. These contributions underscore the significance of \ac{ML} and \ac{DL} in advancing \ac{HEP} research topics.

\subsection{ Related work}
In recent years, there has been a surge in reviews addressing various aspects of \ac{HEP} \cite{abdughani2019supervised,guan2021quantum,stakia2021advances,banerjee2023fifty}. The review presented in \cite{abdughani2019supervised} delved into the realm of supervised \ac{DL} applied to high-energy phenomenology, discussing specific use cases such as employing \ac{ML} to explore new physics parameter spaces and utilizing graph neural networks for particle production and energy measurements at the \ac{LHC}. Meanwhile, paper \cite{guan2021quantum} provided an overview of the initial forays into quantum \ac{ML} in the context of \ac{HEP} and offered insights into potential future applications. In \cite{stakia2021advances}, an array of novel tools relevant to \ac{HEP} were introduced, complete with assessments of their performance, though there was limited discussion about future prospects. Lastly, the review \cite{banerjee2023fifty} comprehensively examined both theoretical and experimental aspects of Jets such as triggering, data acquisition systems, propagation, interactions, and related phenomena in \ac{HEP}. 

Table \ref{tab:rw} assesses how the proposed review aligns with previous research in the field of \ac{HEP}. Based on the assessment, it appears that our proposed review aims to comprehensively cover a wide range of topics related to the ML and DL-based in \ac{HEP}, including Jet preliminaries, taxonomy of \ac{HEP}, available Jet datasets, Jet tagging preprocessing, quantum \ac{ML}, \ac{DL} models for Jet tagging, classification techniques, Jet tagging \ac{DL} applications, and research gaps/future directions. This suggests that the proposed review aims to provide a comprehensive overview of the current state of research in \ac{HEP} and potential avenues for future work.

\begin{table*}[t!]
\caption{Assessing how the proposed review aligns with previous research in the field of \ac{HEP}. The (\cmark) indicates that those specific areas have been addressed, whereas (\xmark) and (\ding{72}) signify instances where certain areas have not been addressed, or partially addressed, respectively. }
\label{tab:rw}
%\scriptsize
%\resizebox{\textwidth}{!}{ 
\begin{tabular}{p{0.8cm}p{1cm}p{0.8cm}p{0.8cm}p{0.8cm}p{0.8cm}p{0.8cm}p{1cm}p{1cm}p{1cm}p{1cm}p{1cm}p{2cm}}
\hline 
\rotatebox{60}{Reference} & \rotatebox{60}{Paper type} & \rotatebox{60}{Publication year} & \rotatebox{60}{\parbox{4cm}{Jet prelimanaries}} &  \rotatebox{60}{Taxonomy of HEP Jet} & \rotatebox{60}{\parbox{4cm}{Available Jet datasets  \newline and tools}} & \rotatebox{60}{Jet tagging pre-process} & \rotatebox{60}{\parbox{4cm}{Quantium ML for HEP\\ 
Jet classification}}  & \rotatebox{60}{\parbox{4cm}{ML and DL models \\ for Jet  classification}} & \rotatebox{60}{\parbox{4cm}{{Transformers  for Jet \\ classification}}} & \rotatebox{60}{\parbox{4cm}{ML and DL-based  Jet \\ classif.  techniques}} & \rotatebox{60}{\parbox{4cm}{AI-based Jet apps}} & \rotatebox{60}{\parbox{4cm}{Research gaps and \\  future direction}}  \\[0.3cm]
 \hline

\cite{abdughani2019supervised} & Mini review & 2019 & \xmark & \xmark & \xmark  & \xmark & \xmark & \xmark & \xmark & \ding{72} & \ding{72} & \xmark \\

\cite{larkoski2020jet} & Review & 2019 & \cmark & \xmark & \xmark & \cmark & \xmark & \xmark & \xmark & \ding{72} & \ding{72} & \xmark \\

\cite{guan2021quantum} & Review & 2021 & \xmark & \xmark & \xmark  & \xmark & \cmark & \xmark & \xmark & \xmark & \xmark & \ding{72} \\

\cite{stakia2021advances} & Review & 2021 & \ding{72} & \xmark & \xmark  & \xmark & \xmark & \ding{72} & \xmark & \ding{72}  & \ding{72}  & \ding{72} \\
\cite{lv2022deep} & Review & 2022 & \cmark & \xmark & \xmark & \ding{72} & \xmark & \ding{72} & \xmark &  \ding{72} & \xmark  & \xmark  \\
\cite{banerjee2023fifty} & Review & 2023 & \cmark  & \xmark & \xmark & \xmark & \xmark & \xmark & \xmark & \xmark & \xmark & \xmark \\
Our & Review & 2024 & \cmark & \cmark & \cmark  & \cmark & \cmark & \cmark & \cmark & \cmark & \cmark & \cmark \\
\hline
\end{tabular}%}
\end{table*}

\subsection{Contribution and survey structure}

The objective of this survey is to provide a robust foundation for both \ac{HEP} researchers aiming to grasp the principles of \ac{DL} and its applications within the \ac{HEP} domain, and for computer science researchers familiar with \ac{AI} seeking insights into the fundamental features and prerequisites essential for constructing a robust \ac{AI} model tailored specifically for \ac{HEP}, employing Jet images {and \ac{PC}}. To achieve this goal, our contribution is encapsulated in the following key points:

\begin{itemize}[leftmargin=6pt] % Cha
    \item The survey offers preliminary insights into the various types of particles and performance metrics associated with both AI-based and non-AI-based Jet particle physics methodologies.

    \item The taxonomy of \ac{ML} and \ac{DL}-based techniques in \ac{HEP} for analyzing Jet images {and \ac{PC}}, along with their respective preprocessing and feature extraction methodologies, is thoroughly explored.
    \item The widely adopted \ac{AI} models designed for analyzing \ac{HEP} Jet {tagging}, along with their descriptive layered architectures, are extensively elaborated upon. Furthermore, their performance metrics are summarized and compared.
    \item Different \ac{SOTA} methods are clustered based on the \ac{AI} techniques employed and comprehensively reviewed accordingly. Additionally, the exploration of AI-based applications in \ac{HEP} Jet {classification} is thoroughly detailed. 
    \item Future directions and outlooks are explored, which aims to offer researchers insights into existing research gaps and areas within \ac{AI} concepts and fields that remain unexplored in AI-based Jet images {and \ac{PC}}.
\end{itemize}

 The structure of this paper is as follows: Section \ref{sec2} presents the preliminaries necessary for understanding Jet images {and \ac{PC}}. In Section \ref{sec3}, the representation of Jet  in DL-based \ac{HEP} is discussed. Section \ref{sec4} provides a summary of the most available \ac{ML} or \ac{DL} models for analyzing \ac{HEP} Jet tagging.  Section \ref{sec6} showcases various AI-based applications of Jet {tagging}. Section \ref{sec7} highlights the gaps and areas that remain unexplored in AI-based Jet analysis, encompassing both techniques and applications. Finally, Section \ref{sec8} concludes the survey.

\begin{figure*}
    \centering
    \includegraphics[scale=1.2]{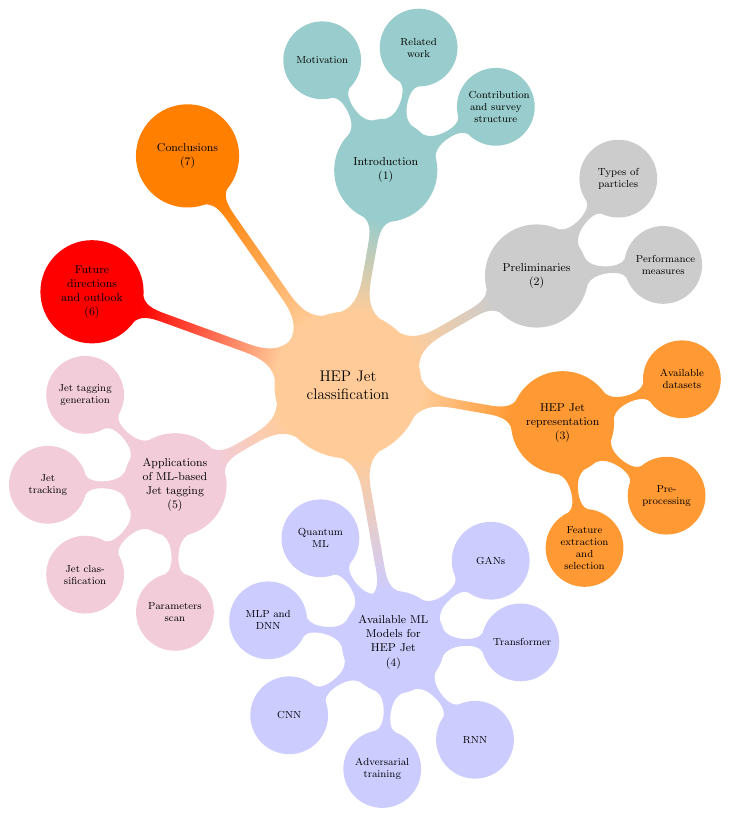}
    \caption{Mind-map of the proposed review.}
    \label{fig:mindMap}
\end{figure*}

\section{Preliminaries}
\label{sec2}

\subsection{Types of particles}

\noindent\textbf{W and Z bosons:} are important closely related particles described by the \ac{SM} of particle physics. They are together known as the weak bosons or more generally as the intermediate vector bosons and plays a significant role in the weak nuclear force, which is responsible for certain types of specific interactions and radioactive decay. The existence and properties of the $Z$ boson, along with the $W$ bosons, provided strong support for the electroweak theory and the \ac{SM} as a whole. However, as with the $W$ boson, the \ac{SM} has limitations and does not explain all aspects of particle physics, such as gravity, dark matter, and the hierarchy of particle masses.  Here are some key points about the $W$ and $Z$ bosons:

\begin{itemize}[leftmargin=6pt] 
\item Charge and Variants: The $W$ boson comes in two varieties: the $W^+$ and the $W^-$, which carry a positive and negative electric charge, respectively. These particles are antiparticles of each other. The $Z$ boson is a neutral elementary particle.

\item Mass and Spin: The $W$ bosons masses are around 80.4 GeV/$c^2$ (gigaelectronvolts per speed of light squared).  The $Z$ boson has a relatively large mass. Its mass is around 91.2 GeV/$c^2$. Both $W$ and $Z$ bosons have a spin of 1, which is a measure of their intrinsic angular momentum. 

\item Decay: The $W$ and $Z$ bosons are unstable and have a very short lifetime. They quickly decay into other particles. For example, a $W^+$ boson can decay into a positron (an antielectron) and a neutrino, while a $W^-$ boson can decay into an electron and an antineutrino. The $Z$ can decay into various combinations of charged leptons (such as electrons and muons) and their corresponding antiparticles, as well as neutrinos and antineutrinos.
\end{itemize}

\noindent\textbf{The Higgs boson:} is crucial to our understanding of how other particles acquire mass and, by extension, how the universe's structure and behavior arise. The key points about the Higgs boson are \cite{jakobs2024profile}:

\begin{itemize}[leftmargin=6pt] 
    \item \textbf{Origin of Mass:} is associated with the Higgs field, a theoretical field that permeates all of space. In the \ac{SM}, particles acquire mass by interacting with the Higgs field. The more a particle interacts with this field, the greater its mass will be. This mechanism explains why some particles are heavier than others.
    \item \textbf{Mass and Spin:} the Higgs boson itself has a mass of around 125.1 GeV/$c^2$ . It has a spin of 0, which means it has no intrinsic angular momentum.
    \item \textbf{Decay:}  is unstable and quickly decays into other particles after its creation in high-energy collisions. The specific decay modes and products depend on the energy at which it is produced.
    \item \textbf{Higgs field Interaction:}  is a carrier of the interaction associated with the Higgs field. When particles move through space, they interact with this field, which gives them mass. The Higgs boson itself is the quantized excitation of this field.
    
\end{itemize}

\noindent\textbf{The top quark:} is one of the heavy fundamental particles described by the \ac{SM}. It holds a special place in particle physics due to its extremely large mass and its role in various processes involving high-energy collisions. Here are some key points about the top quark \cite{hoang2020top}:

\begin{itemize}[leftmargin=6pt] 
    \item \textbf{Mass:} The top quark is the heaviest known elementary particle. Its mass is approximately 173.2 GeV/$c^2$, which is even heavier than an entire atom of gold.
    \item \textbf{Quarks and the strong force}: Quarks are the building blocks of protons and neutrons, which are the constituents of atomic nuclei. The top quark, like all quarks, experiences a strong nuclear force, which is responsible for holding quarks together within hadrons (particles composed of quarks).
    \item \textbf{Weak decays:} Due to its high mass, the top quark is relatively short-lived and decays before it can form bound states with other quarks to create hadrons. It decays primarily through weak interaction, one of the fundamental forces described by the \ac{SM}.
    \item \textbf{Production and detection:} The top quark is typically produced in high-energy particle collisions, such as those that occur in experiments at particle accelerators like the \ac{LHC}. Due to its high mass, the top quark is often produced along with its corresponding antiquark. Researchers detect its presence indirectly by observing its decay products, which can include other quarks, leptons (such as electrons and muons), and neutrinos.

    \item \textbf{Role in electroweak symmetry breaking:} The top quark is of particular interest in theories related to electroweak symmetry breaking, a phenomenon that explains why certain particles acquire mass. Its large mass plays a significant role in the behavior of the Higgs boson and its interactions.
\end{itemize}

\noindent\textbf{The $b$ and $\bar{b}$ Jets:}

Jets composed of $b$ and $\bar{b}$ pairs are identified by mandating a minimum \ac{$p_T$} of $20 \, \text{GeV}/c$ for each Jet and restricting their pseudorapidity ($\eta$) to the interval $2.2 < \eta < 4.2$. This criterion ensures the Jets are well contained within the detector's instrumented region. Following initial selection, 16 distinct Jet substructure features are utilized as inputs for the classification algorithms. Within a Jet, the highest \ac{$p_T$} muon, kaon, pion, electron, and proton are chosen. For each of these particles, three physical parameters are evaluated: the relative transverse momentum to the Jet's axis ($p_{rel}^T$), the electric charge ($q$), and the separation in the $(\eta,\phi)$ space from the Jet axis ($\Delta R$). Should any particle type be absent, its corresponding features are assigned a value of 0. An additional characteristic, the weighted Jet charge $Q$, is computed as the sum of the particles' charges inside the Jet, each multiplied by its respective $p_{rel}^T$ \cite{gianelle2022quantum}.

\begin{figure}
    \centering
    \includegraphics[scale=1]{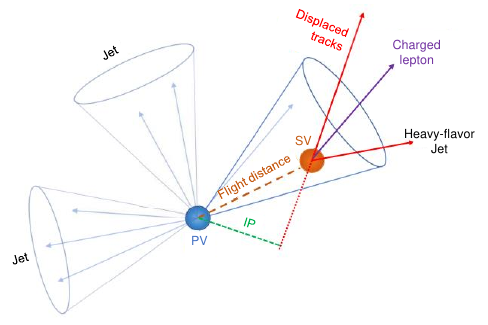}
    \caption{Visualization of decay involving a reconstructed Jet and a secondary vertex, showcasing various noteworthy features \cite{novak2020sissa}.}
    \label{fig:jet-param}
\end{figure}

\subsection{Key concepts of ML-based HEP}

When discussing \ac{ML} and its subset \ac{DL} in \ac{HEP}, maintaining uniform and precise terminology is crucial for clear communication. \textit{Supervised learning}, for instance, refers to training models using labeled datasets, where the model learns to map input features to known outputs, such as identifying particles or classifying Jets based on their physical properties. In contrast, \textit{unsupervised learning} involves identifying patterns or structures in data without predefined labels, often used in anomaly detection or clustering in particle physics. \textit{Feature selection} is an essential process that focuses on choosing the most informative input features—such as track momentum, calorimeter energy deposits, and hit patterns in detectors—thereby improving the performance and efficiency of \ac{ML} models by reducing dimensionality and computational load. 

The growing adoption of \ac{DL} techniques, like  
 \acp{CNN} and \acp{GNN}, has revolutionized analyses in \ac{HEP}. These methods rely on different types of layers and architectures designed to handle the complexity and scale of particle physics data. \textit{Convolutional layers} in \acp{CNN}, for instance, are particularly effective at detecting patterns in images or {\acp{PC}}, by learning local features. These layers operate by applying convolutional filters to input data, extracting hierarchical patterns, which are then pooled to reduce dimensionality. Pooling layers, such as \textit{max-pooling}, downsample the spatial dimensions of the data, retaining the most important features while reducing computational cost. This structure allows \acp{CNN} to efficiently process large-scale data and is widely used in Jet classification and particle identification tasks. Further advancements include the use of \textit{EdgeConv layers} in \acp{GNN} \cite{semlani2024pcn}, where the network learns the relationships between particles represented as nodes in a graph. In these models, the \textit{EdgeConv} block aggregates local particle information, capturing spatial relationships and interactions based on particle kinematics and connectivity, which are essential for Jet tagging. The use of \textit{global average pooling} in these models helps aggregate information from individual particles, producing a global representation of the Jet that can then be used for classification or regression tasks. \textit{Dense layers} (also known as fully connected layers) play a critical role in transforming high-level features learned by convolutional and graph-based layers into a final prediction. Dense layers are used in \acp{DNN}, \acp{CNN}, \acp{GNN}, and others, after the feature extraction phase, where the output of the convolutional or graph layers is flattened into a one-dimensional vector and passed through one or more fully connected layers. These layers allow the network to combine the learned features in a non-linear way, making complex decisions such as event classification, particle identification, or regression for Jet properties. The dense layer's ability to connect all input neurons to all output neurons allows the model to capture intricate relationships between features, making it highly effective for tasks like anomaly detection, signal classification, and event reconstruction in \ac{HEP}.

An essential innovation in modern \ac{DL} is the \textit{Attention layer} \cite{vaswani2017attention}, a core layer in building Transformers, enables the model to focus on the most relevant parts of the input data. Attention mechanisms are particularly useful in scenarios where certain elements in a sequence (or graph) are more important for the task than others. In particle physics, this could involve focusing on particular particle interactions or energy deposits in Jets. The \textit{scaled dot-product Attention} mechanism, used in Transformer models, computes attention scores for each pair of input elements \cite{kheddar2024automatic}. The attention output \( \text{Attention}(Q, K, V) \) is calculated as follows:

\begin{equation}
    \text{Attention}(Q, K, V) = \text{softmax}\left(\frac{QK^T}{\sqrt{d_k}}\right) V
\end{equation}

Where \( Q \), \( K \), and \( V \) represent the \textit{query}, \textit{key}, and \textit{value}  matrices, respectively, and \( d_k \) is the dimension of the key vectors. The softmax function normalizes the attention scores, allowing the model to weigh the importance of different elements in the input sequence. This mechanism enables the model to prioritize relevant information, improving the accuracy of particle event classification, Jet tagging, and anomaly detection, particularly when the input data has complex dependencies or long-range interactions between particles.

\color{black}

\subsection{Performance measures}

In the realm of \ac{HEP}, performance assessment is divided into two main categories. The first encompasses classical metrics like energy loss, path length, and axis distance. The second involves metrics related to \ac{DL}-based \ac{HEP} techniques, such as accuracy, \ac{TPR}, \ac{FPR}, \ac{ROC}, \ac{AUC}, \ac{MSE}, \ac{FST}, among others. Table \ref{tab:metrics} outlines these metrics, including mathematical formulations and descriptions.

\renewcommand{\arraystretch}{1.5} % Increase row height by 1.5 times

\begin{table*}
\centering
\scriptsize
\caption{An overview of the metrics employed to evaluate performance in  {ML and DL-based HEP.}}\label{tab:metrics}
\begin{tabular}{m{10mm}m{30mm}m{10mm}m{100mm}}
 \\
\hline
 Metric & Formula & {C/R}  & Description   \\ 
\hline \hline
\Ac{FPR} and \Ac{TPR} & \(\displaystyle \mathrm{\frac{FP}{FP+TN}}, \mathrm{\frac{TP}{TP+FN} }\) & {C} &  The \Ac{FPR}, is the ratio (or percentage)  of the background signal that are incorrectly identified as containing Jet.  The \Ac{TPR}. is the ratio (or percentage) of the Jet signal that is correctly identified as Jet (particle). \\ \hline
\Ac{AUC}& \(\displaystyle
\int_{0}^{1}  \mathrm{TPR} \: d(\mathrm{FPR}) \)& {C} &   The area beneath the \ac{ROC} curve is represented. It delivers a singular numeric score reflecting the cumulative effectiveness of the classification technique. An elevated \ac{AUC} score signifies superior performance, with the ideal score being 1.\\\hline

Accuracy & \(\displaystyle \mathrm{\frac{TP+TN}{TN+FN+TP+ FP}} \) & {C} &  The accuracy is the ratio (or percentage) of correctly detected instances of Jet in the signal. A high accuracy indicates that the classification algorithm is more effective in detecting Jet than background.\\\hline
\Ac{MSE} & $\frac{1}{N}\sum_{i=1}^N (P_b^i(\theta)-T^i)^2$ &  {R} &   The training procedure seeks to discover the model parameter values denoted as $\theta$, which minimize the loss function known as \ac{MSE}. Where $N$ is the number of training Jets, $P_b^i$ and $T^i$ is the predicted and target probabilities, respectively, for the $i$-th Jet. \\ \hline

F1-score& \(\displaystyle 2. \mathrm{\frac{Precision \times Recall}{ Precision+Recall}} \) & {C} &   Represents the harmonic mean between precision and recall metrics. This measure is applied to assess the comprehensive efficacy of the classification algorithm in identifying or tagging Jets. \\
\bottomrule
\end{tabular}
\begin{flushleft}
 Abbreviations:  Classification or regression (C/R);   
\end{flushleft}
\end{table*}

\section{HEP Jet representation}
\label{sec3}

This section provides an overview of the Jet datasets comprising various forms of Jet data obtained and generated through different methods. Additionally, the current section delves into diverse pre-processing and feature extraction techniques employed in this context.

\subsection{Available datasets {and simulation tools}}
\begin{figure*}
    \centering
    \includegraphics[scale=1.1]{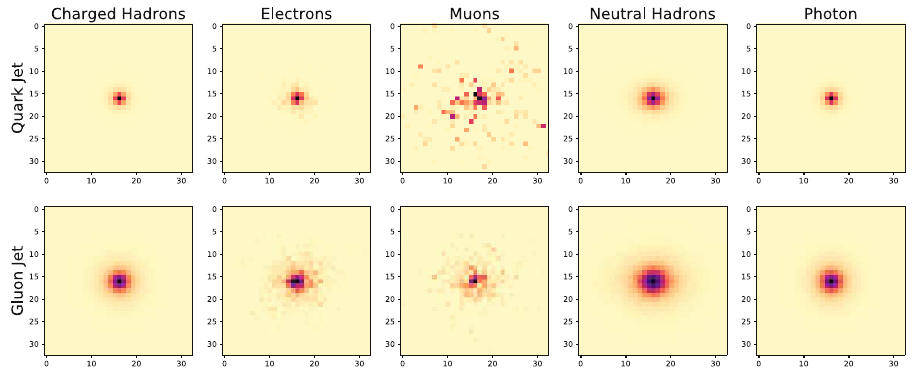}
    \caption{Jet images summed online and categorized into different channels employed in the analysis within the 100-200 GeV \ac{$p_T$} range.
}
    \label{fig:quark-gluon}
\end{figure*}

The \ac{CERN} open data portal provides access to a variety of datasets from experiments conducted at the Large \ac{LHC}. These datasets include information about collisions, particles, and Jet images {and \acp{PC}}. The portal offers a great starting point for those interested in \ac{HEP} datasets.  Figure \ref{fig:quark-gluon} illustrates samples of Jet images, featuring the average of \ac{$p_T$}-normalized quark and gluon Jet images across 5 distinct $\chi$ bins. The  Jet images {or \ac{PC}} may  undergo different preprocessing techniques, discussed later, prior to input into ML/DL models for classification or prediction tasks. Table \ref{tab:datasets} presents the datasets, {along with several simulation tools}, most commonly used in the research reviewed in this paper.

\begin{table*}
\scriptsize
\caption{A summary of available datasets, and {simulation tools} for Jet \ac{HEP} analysis. }
\begin{tabular}[!t]{m{5mm}m{20mm}m{130mm}m{10mm}}

\label{tab:datasets} \\
\toprule
 & Name & Description & DLA ?  \\ 
\hline \hline
\multirow{7}{*}{\rotatebox{90}{Datasets}} & ATLAS open data &  Is one of the largest particle physics experiments at the \ac{LHC}. They offer an "Open Data" initiative with datasets that include collision data and simulated samples. These datasets can be used to study Jet images and other particle physics phenomena. & {Yes}\tablefootnote{\url{http://opendata.cern.ch/search?page=1&size=20&experiment=ATLAS}}\\ 

 & CMS open data &  \Ac{CMS} is another major experiment at the \ac{LHC}. Similar to ATLAS, \ac{CMS} provides open data for educational and research purposes. The datasets include information about collisions, particles, and Jets. & {Yes} \tablefootnote{\url{https://opendata.cern.ch/search?page=1&size=20&q=jet\%20images&experiment=CMS}} \\ 
 
 & Complete  &  It belongs to \ac{CERN} and contains muon, kaon, pion, electron, and proton. In the complete dataset training, 400,000 Jets are used for training, and the remaining 290,000 are used for testing and assessing performance \cite{gianelle2022quantum}. & No \\ 

  & Top tagging & This dataset comprises 1.2 million training samples, 400,000 for validation, and another 400,000 for testing. Each entry in this dataset corresponds to an individual Jet, with its source being either an energetic top quark, a light quark, or a gluon. These events were generated using the PYTHIA8 Monte Carlo event generator, and the response of the ATLAS detector is simulated using the DELPHES software package & {Yes}\tablefootnote{\url{https://zenodo.org/record/2603256}} \\ 

 & Quark-gluon tagging & The dataset is created by generating signal (quark) and background (gluon) Jets through PYTHIA8. For the signal Jets, the process involves \(Z( \rightarrow νν) + (u, d, s)\), and for the background Jets, it uses $Z(\rightarrow νν) + g$. Notably, there is no simulation of the detector. The particles that are not neutrinos in the final state are grouped into Jets using the anti-kT algorithm with a radius parameter of $R = 0.4$. In total, this dataset contains 2 million Jets, evenly split between signal and background categories \cite{gong2022efficient}. & No\\
 & Higgs dataset & The dataset originates from Monte Carlo simulations. The initial 21 attributes (found in columns 2-22) represent particle detector-derived kinematic properties within the accelerator. The remaining seven attributes are transformations of the initial 21, constituting high-level features engineered by physicists to aid in distinguishing between the two categories. & Yes\tablefootnote{\url{ https://archive.ics.uci.edu/dataset/280/higgs}}\\

 & QCD multi-Jet &
Samples are generated across different ranges of scalar sum of \ac{$p_T$}, namely 1000-1500 GeV, 1500-2000 GeV, and 2000-Inf GeV. After excluding samples with \ac{$p_T$} values less than 1000 GeV, the dataset consists of around 450. 10$^3$ training images, 150.10$^3$ validation images, and 150.10$^3$ testing images \cite{kim2023multi}. & No\\
\hline 
\multirow{7}{*}{\rotatebox{90}{Simulation tools}} & Delphes  &  Is a particle physics event generator designed to produce simulated collision events that are similar to those observed in real experiments. It includes tools to generate Jet  based on the data produced in simulations. & {Yes}\tablefootnote{\url{https://cp3.irmp.ucl.ac.be/projects/delphes}} \\ 
 & MadGraph &  Is a popular event generator used in particle physics simulations. It can generate events involving Jets and other particles, which can then be turned into Jet {\ac{PC} or} images. & {Yes}\tablefootnote{\url{http://madgraph.phys.ucl.ac.be/}} \\ 
 & FASTSim &  Is a tool for simulating high-energy particle collisions. It can generate Jets from simulated collision events and is often used for studying \ac{ML} techniques in \ac{HEP}. & {Yes}\tablefootnote{\url{https://twiki.cern.ch/twiki/bin/view/CMSPublic/SWGuideFastSimulation}} \\ 

 & Monte Carlo &  It is generated through a dependable framework, created by integrating various tools like Pythia 8 for generating \ac{HEP} events, Delphes for emulating the detector's response, and RAVE for reconstructing secondary vertices \cite{di2023automated}. & No \\
\hline
\bottomrule
\end{tabular}
\begin{flushleft}
Abbreviations: \ac{DLA}. 
\end{flushleft}
\end{table*}

\subsection{Pre-processing for ML-based Jet  analysis}
The objective of preprocessing input data is to support the model in addressing an optimization challenge. Usually, these preprocessing actions are not mandatory, but they are employed to enhance the numerical convergence of the model, considering the real-world constraints imposed by limited datasets and model dimensions, along with the specific parameter initialization choices. In \ac{HEP},  (i) $\eta$  represents pseudorapidity, which is a measure related to the polar angle of a particle's trajectory. It is commonly used because it is less affected by relativistic effects and is approximately invariant under boosts along the beamline, (ii)  $\phi$ represents the azimuthal angle, which is the angle around the beamline, (iii) together, $\eta$ and $\phi$ provide a way to specify the direction and position of particles or energy deposits within the detector. These coordinates are particularly useful for representing and analyzing the distribution of particles produced in high-energy collisions, (iv) the combination of $\eta$ and $\phi$ can be thought of as a way to navigate and map the detector's components in a way that is sensitive to the underlying physics processes,  (v)
$\eta - \phi$ space is a coordinate system used to describe the properties and positions of particles or objects within particle detectors, particularly in experiments at large colliders like the \ac{LHC}.

The subsequent sequence of data-driven preprocessing procedures was employed on the Jet images and can also be adapted for \acp{PC}):
\begin{itemize}[leftmargin=6pt] 
    \item\textbf{Center (translation and  rotation):} Center the Jet image by translating it in $(\eta, \phi)$ coordinates, such that the pixel with the centroid weighted by total \ac{$p_T$} is located at $(\eta, \phi) = (0,0)$. This procedure involves rotating and boosting the Jet along the beam direction to position it at the center.

    \item\textbf{Crop:} Trim to a region of $value \times value$ pixels centered around $(\eta, \phi) = (0,0)$, encompassing the area where $\eta, \phi$ fall within the range $(-R, R)$.

    \item\textbf{Normalize:} adjust the pixel intensities to ensure that the sum of all pixel values, $\sum_{i,j} I_{i,j}$, equals 1 across the image, with $i$ and $j$ serving as the pixel indices. 
     \item\textbf{Zero-center:} Remove the average value, represented by $\mu_{i,j}$, from the normalized training set images from every image, thereby altering each pixel's intensity to $I_{i,j} = I_{i,j} - \mu_{i,j}$.
   \item\textbf{Standardize:} Normalize each pixel by dividing it by $\sigma_{i,j}$ (the standard deviation) of the corresponding pixel value in the training dataset. This process is represented as: $I_{i,j} = I_{i,j}/ ( \sigma_{i,j}+r)$. A value of $r = 10^{-5}$ was employed to reduce the influence of noise.
   \item\textbf{Clustering and trimming:} Reconstruct Jets by applying the anti-$k_t$ algorithm \cite{cacciari2008anti} to all calorimeter towers, utilizing a specific Jet size parameter, such as $R = 1.0$, and then choose the primary (leading) Jet. Subsequently, refine the Jet by employing the $k_t$ algorithm with a subjet size parameter of $r < R$, such as $r = 0.3$  \cite{barnard2017parton}.
   \item \textbf{Pixelisation:} Create a Jet image by discretizing the transverse energy of the Jet into pixels with dimensions (0.1, 0.1) in the $\eta - \phi$ space.
   \item \textbf{Zooming:} It is  the option to magnify the Jet image by a factor that diminishes its reliance on the Jet's momentum.
\end{itemize}

\subsection{Feature extraction and selection}

Feature extraction and selection are important techniques in \ac{HEP} for analyzing and interpreting data from experiments conducted at particle accelerators like the \ac{LHC}. \ac{HEP} experiments produce vast amounts of data, and the goal is to extract relevant characteristics from this data to make: (i) particles identifications, (ii) extract kinematic variables, such as \ac{$p_T$}, energy (E), rapidity (y), and azimuthal angle ($\phi$) for each detected particle, (iii) calculating the invariant mass of particle can reveal the presence of new particles, (iv)  extract topological features related to the spatial distribution of particles or their interactions such as angular separations, impact parameters, and vertex finding. The benefit of feature selection is to make: (i) dimensionality reduction techniques like \ac{PCA} or \ac{t-SNE} may be employed to reduce the number of features while retaining as much information as possible, (ii)  identify the most discriminating features that separate signal from background, (iii) identify the most relevant features for \ac{ML} classification and model building.

Di Luca et al. \cite{di2023automated} presents an automated feature selection procedure for particle Jet classification in \ac{HEP} experiments. The authors use \ac{ML} boosted tree algorithms to rank the importance of observables and select the most important features associated with a particle Jet. They apply this method to the specific case of boosted Higgs boson decaying to two
b-quarks ($H \rightarrow bb$) tagging and demonstrate the impact of feature selection on the performance of the classifier to distinguish these events amidst the substantial and unalterable background originating from \ac{QCD} multi-Jet production. They also train a fully connected neural network to tag the Jets and compare the results obtained using all the features or only those selected from the procedure which consists of two main steps: data preparation and feature ranking extraction. The authors discover that the \textit{azimuthal angles} of the large-R Jet and the \textit{variable radius (VR)-track} Jets appear towards the end of the feature ranking. At the top of the ranking, they find the \ac{$p_T$} of the two VR-track Jets, along with certain details regarding the secondary vertex, such as its \textit{mass, energy}, and \textit{displacement}. The study shows that selecting the highest-ranked features achieves performance nearly as effective as that of the full model, with only a slight deviation of a few percent. This approach can be expanded to accommodate the increased number of observable variables that upcoming collider experiments will gather from high \ac{$p_T$} particle Jets. The data for this research comes from proton-proton collision events featuring a boosted Higgs boson that decays into two $b$ quarks.  In \cite{strong2020impact}, solutions have been proposed for classifying events extracted from the 2014 Higgs ML Kaggle dataset\footnote{\url{https://www.kaggle.com/c/higgs-boson}}. The dataset includes a mix of low-level and high-level attributes: it contains 18 low-level features that include three-dimensional momenta (\(p_T\), \(\eta\), \(\phi\)), missing transverse momentum, and the total transverse momentum from all Jets; additionally, there are 13 high-level features motivated by physics, covering invariant masses and angular separations among objects in the final state. Table \ref{tab:features} summarizes the features utilized, which hold potential for future application within the context of \ac{HEP}.  The authors aim to ensure that the suggested networks make effective use of low-level information; otherwise, there's a risk of losing these features during selection. Their focus lies in determining the necessity of high-level features. The proposed \Ac{DNN} model effectively utilize the low-level information in the data and autonomously learn their own high-level representations.  \Ac{BIP} features are a type of mathematical representation used in \ac{HEP} for analyzing particle collision data. They are constructed to be invariant under boosts, meaning they remain unchanged under transformations to different reference frames with different velocities. These features are designed to capture important characteristics of particle Jets, such as their energy distribution and substructure, while ensuring consistency across various experimental conditions. \ac{BIP} features are particularly useful for tasks like Jet tagging and classification in \ac{HEP} experiments, as employed in \cite{munoz2022boost}.

\begin{table*}
\scriptsize
\caption{Possible combinations of Jet features to generate new high- and low-level features that could potentially improve \ac{ML} classification for Jet \ac{HEP}. The performance of employing these features are presented in \cite{strong2020impact}. }
\begin{tabular}[!t]{m{3mm}m{40mm}m{90mm}m{30mm}}

\label{tab:features} \\
\toprule
Level & Suggested feature name & Description  & Grouping  \\ 
\hline \hline
& DER\_mass\_MMC & The Higgs boson's mass was estimated using a hypothesis-driven fitting method & Higgs, Mass\\
& DER\_mass\_transverse\_met\_lep  & Transverse mass associated with the lepton and \(P_{\text{miss}}^{T}\) & Higgs, Mass \\
& DER\_mass\_vis  & The mass invariant to both the lepton and the tau & Higgs, Mass\\
\multirow{2}{*}{\rotatebox{90}{High-level features}}  & DER\_pt\_h  & Transverse momenta of the combined vector of the lepton, tau, and $P_{\text{miss}}^{T}$
 & Higgs, 3-momenta\\
  & DER\_deltaeta\_jet\_jet  & Absolute disparity in pseudorapidity between the leading and subleading Jets (undefined for less than two Jets)
 & Jet with angular \newline properties\\
  & DER\_mass\_jet\_jet & The invariant mass of the primary and secondary Jets (not applicable when there are fewer than two jets)
 & Jet, Mass \\
  & DER\_prodeta\_jet\_jet & The multiplication of the pseudo rapidities for the foremost and next-to-foremost Jets (inapplicable if fewer than two Jets are present)  & Jet, 3-momenta \\
  & DER\_deltar\_tau\_lep  & Distance between the lepton and the tau in the $\eta - \phi$ plane
 & Final state, Angular \\
  & DER\_pt\_tot & The \ac{$p_T$} resulting from the vector addition of the \ac{$p_T$} of the lepton, tau, the primary and secondary Jets (when applicable), and $P_{\text{miss}}^{T}$
 & Final-state, Sum \\
 & DER\_sum\_pt  & Total transverse momentum of the lepton, tau, and all Jets
 & global event, Sum \\
 & DER\_pt\_ratio\_lep\_tau  & Ratio of the transverse momenta of the lepton to that of the tau
& Final state, 3-momenta \\
  & DER\_met\_phi\_centrality  & Centrality of the azimuthal angle of $P_{\text{miss}}^{T}$ relative to the lepton and the tau
 & Final state, Angular \\
  & DER\_lep\_eta\_centrality  & The centrality measure of the lepton's pseud-orapidity in comparison to the primary and secondary Jets (not applicable for fewer than two Jets)
 & Jet, Angular \\  \hline
  & PRI\_tau\_[px/py/pz]  & The 3-momenta of the tau expressed in Cartesian coordinates & Final state,  3-momenta \\
 \multirow{2}{*}{\rotatebox{90}{Low-level features}}  & PRI\_lep\_[px/py/pz]  & The lepton's 3-momenta represented in Cartesian coordinates & Final state, 3-momenta\\
  & PRI\_met\_[px/py]  & The constituent parts of the missing transverse momentum vector expressed in Cartesian coordinates & Final state, 3-momenta\\
  & PRI\_met  & The magnitude of the missing transverse momentum vector represented in Cartesian coordinates & Final state, 3-momenta\\
  & PRI\_met\_sumet  & Total sum of transverse energy & Final-state, Energy\\
  & PRI\_jet\_num  & Count of Jets present in the event & Jet, Multiplicity \\
  & PRI\_jet\_leading\_[px/py/pz]  & The three-dimensional momenta of the primary Jet expressed in Cartesian coordinates (not applicable if there are no Jets present) & Jet, 3-momenta \\
  & PRI\_jet\_subleading\_[px/py/pz]  & The 3-momenta of the secondary Jet represented in Cartesian coordinates (not defined if fewer than two Jets are present) & Jet, 3-momenta\\
  & PRI\_jet\_all\_pt  & Total sum of the transverse momenta of all Jets in Cartesian coordinates & Jet, 3-momenta\\[2mm]
\bottomrule
\end{tabular}
\begin{flushleft}
Note: PRI\_jet\_all\_pt may diverge from the sum of the transverse momenta of the leading and subleading jets because events can feature more than two jets. 
\end{flushleft}
\end{table*}

%\cite{ma2017novel}

\section{Available AI Models for HEP  {Jet classification}}
\label{sec4}

Many \ac{DL} architectures have been proposed in the \ac{SOTA} of \ac{HEP} domain to identify particles. {Some of these architectures require input data in the form of images, while others utilize \ac{PC} representations  \cite{sohail2024advancing}.} Table \ref{tab:archCmp} summarizes and compares the most efficient \ac{ML} and \ac{DL} models, used in \ac{HEP}, based on their architectures and performances.

\begin{figure*}
    \centering
    \includegraphics[scale=0.9]{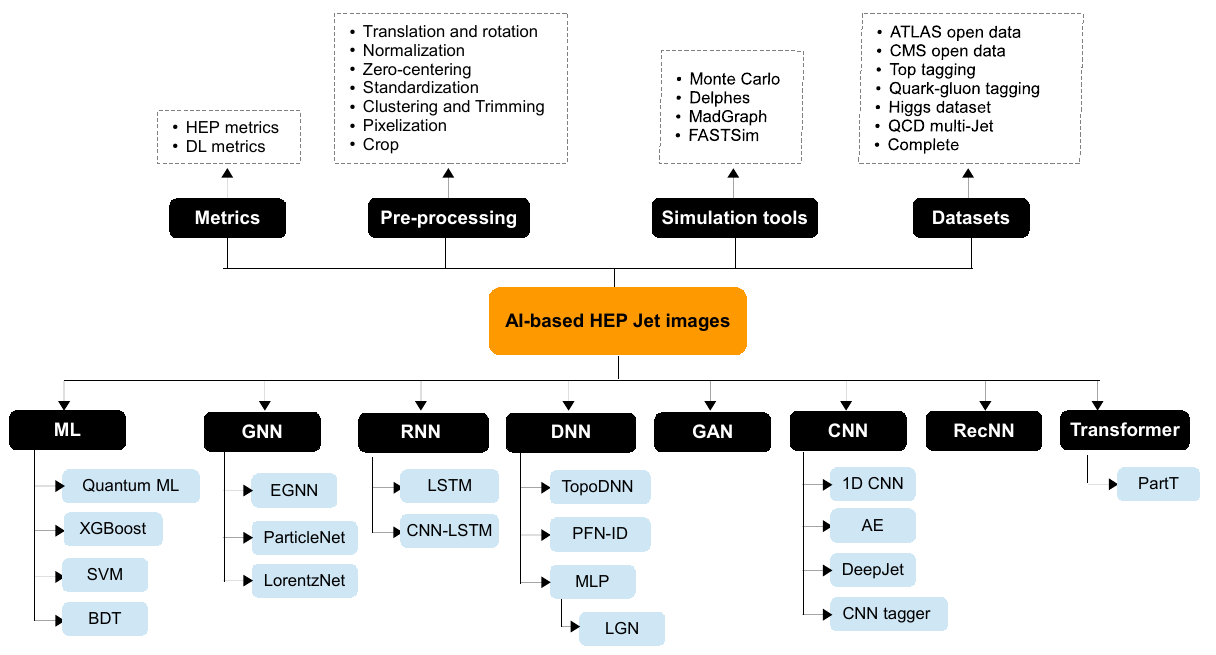}
    \caption{Taxonomy of ML and DL-based HEP techniques for Jet {classification}, with associated preprocessing, metrics, simulation tools and datasets. }
    \label{fig:TaXmethods}
\end{figure*}

\Ac{ML}, especially \ac{DL}, has a rich historical presence in the field of particle physics. The concept of applying neural networks for tasks like distinguishing quarks and gluons, tagging Higgs particles, and identifying particle tracks has been around for more than two and a half decades. Nevertheless, the recent advancements in \ac{DL} and the increased computational capabilities offered by \acp{GPU} have led to a significant enhancement in image recognition technology. As a result, there has been a renewed and heightened interest in utilizing these techniques. In the subsequent sections, we provide an overview of \ac{SOTA} methods in both \ac{ML} and \ac{DL}. Figure \ref{fig:TaXmethods} depicts  a taxonomy of existing \ac{ML} and \ac{DL} techniques, summarizes the reviewed AI-based Jet classification models (discussed in Section \ref{sec4}), preprocessing and  datasets (discussed in Section \ref{sec3}), and metrics (discussed in Section \ref{sec2}).

\subsection{ML-based methods} \label{sec5.1}
\ac{ML}-based analysis of \ac{HEP} Jet tagging has become an important technique in recent years. Jets are collimated sprays of particles, i.e, emitted from a source in a way that they are parallel or nearly parallel to each other, produced in high-energy particle collisions. Analyzing their properties is crucial for understanding the underlying physics processes. Jet images and \ac{PC} are essentially 2D and 3D representations of the energy distribution within a Jet, where each pixel corresponds to a small region of the Jet. For example, 
 Bogatskiy et al. in \cite{bogatskiy2022pelican} introduced PELICAN, an \ac{ML} architecture for particle physics that leveraged permutation-equivariant and Lorentz-invariant techniques, along with elementary equivariant aggregators and dense message-passing blocks. It processed 4-vector inputs representing particle jets as point clouds and employed a classifier to reduce rank-2 input arrays (pairwise dot products of 4-momentum vectors of particles in a jet) to permutation-invariant scalars using trace and total sum aggregation functions. Dense layers and a cross-entropy loss function were then used for optimization. Additionally, the PELICAN regressor predicted 4-momentum of particles using a permutation- and Lorentz-equivariant architecture with rank-preserving transformations and loss functions based on relative momentum and mass resolutions. Evaluation metrics included accuracy, \ac{AUC}, background rejection rate, and relative resolutions. PELICAN achieved state-of-the-art performance in Jet classification, outperforming methods like LorentzNet while using approximately five times fewer parameters (45k only). Its low complexity, enhanced by equivariant aggregation, message-passing mechanisms, and its ability to handle regression tasks, made it suitable for real-time applications. However, its limitations included evaluation on limited datasets and reliance on hyperparameter tuning.

\color{black}

\begin{figure*}
    \centering
    \includegraphics[scale=0.8]{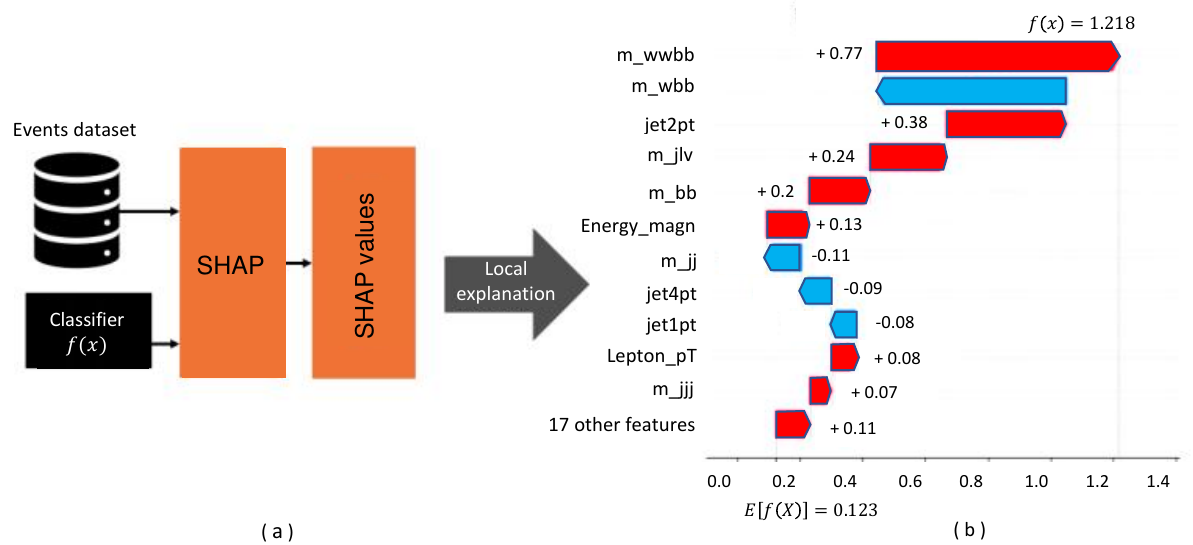}
    \caption{(a) Diagram illustrating the localized explanation of an event classifier with the \ac{SHAP} method. (b) Localized \ac{SHAP} explanation represented using a waterfall plot. It can be observed that the \ac{SHAP} values are associated with individual event features. The classifier's prediction (XGBoost) is $f(x) = 1.218$, while the base value is $E[f(x)] = 0.123$. In this context, the feature "m\_wwbb" contributes positively with a \ac{SHAP} value of +0.77, increasing the prediction, whereas the feature "m\_wbb" has a \ac{SHAP} value of -0.6, reducing the prediction.}
    \label{fig:shap}
\end{figure*}

\ac{ML} technique have been used in  \cite{pezoa2023explainability} by applying the \ac{SHAP} method to explain the output of two \ac{HEP} events \ac{ML} classifiers (XGBoost and \ac{DNN}) using the Higgs dataset. It demonstrates \ac{SHAP}'s utility in understanding complex \ac{ML} systems, particularly in the context of \ac{HEP} event classifiers. The TreeExplainer and DeepExplainer methods from the Python \ac{SHAP} library were used to compute \ac{SHAP} values, revealing that features like $m\_bb$, $m\_wwbb$, and $m\_wbb$ were crucial in both models, although their distribution of \ac{SHAP} values differed, indicating distinct learning processes. The process of extracting \ac{SHAP}  values are depicted in Figure \ref{fig:shap}.

In addition, \ac{QML} methods have recently found applications in addressing challenges within \ac{HEP}, including separating signal from background \cite{blance2021quantum}, detecting anomalies \cite{blance2021unsupervised}, and reconstructing particle tracks \cite{funcke2023studying}.

 Blance and Spannowsky \cite{blance2021quantum} proposed a hybrid variational quantum classifier that combines quantum computing methods with classical neural network techniques to improve classification performance in particle physics research. The algorithm is applied to a resonance search in di-top final states, and it outperforms both classical neural networks and \ac{QML} methods trained with non-quantum optimization methods. The classifier's ability to be trained on small amounts of data indicates its potential benefits in data-driven classification problems. The proposed methodology was applied to the generated dataset, and the hybrid approach using the \ac{FST} metric outperformed both classical neural networks and \ac{QML} methods trained with non-quantum optimization methods in terms of maximizing learning outcomes; its accuracy can reach 72.6\%. The hybrid approach also learned faster than an equivalent classical neural network or the classically trained variational quantum classifier. The paper \cite{sharma2021quantum} discusses the potential applications of quantum computation and \ac{QML} in \ac{HEP}, rather than focusing on deep mathematical structures. The authors claim that statistical \ac{ML} methods are used for track and vertex reconstruction. These methods vary depending on the detector geometry and magnetic field used in the experiment. \ac{ML} can help address these challenges by providing efficient and accurate methods for pattern recognition and particle identification. They suggest that quantum algorithms could potentially improve upon existing methods by offering faster and more efficient solutions to challenging problems in experimental \ac{HEP}, such as particle identification and track reconstruction. This can be realized by creating a dataset recorded on tape through grid computing, which can be distributed for offline analysis using \ac{QML} to extract information about particle trajectories developed inside the detectors. The work \cite{wu2021application} investigates the potential of \ac{QML} in \ac{HEP} analysis at the \ac{LHC}. The authors compare the performance of the quantum kernel algorithm to classical \ac{ML} algorithms using 15 input variables and up to 50,000 events. They used 60 statistically independent datasets of 20,000 events each for their analysis. The \ac{AUC} is used as the metric, and the results show that the performance of all methods improves with increasing dataset size. For 15 qubits, the quantum SVM-Kernel algorithm performs similarly to the classical \ac{SVM} and classical \ac{BDT} algorithms. The quantum SVM-Kernel performances from the three different quantum computer simulators (Google, IBM, and Amazon) are comparable. The authors also claim that when a selection is implemented, permitting a signal acceptance rate of 70\%, it results in the rejection of approximately 92\% of background events, as indicated by the \ac{AUC}. Consequently, the $S/\sqrt{B}$ ratio will experience an enhancement of approximately 150\% compared to a scenario without any selection. Similarly, the researchers in \cite{gianelle2022quantum} present a new approach to Jet classification using \ac{QML}. The method involves embedding data into a quantum state, passing it through a variational quantum circuit, and performing a training procedure by minimizing a classical loss function. Probability measurements of the final state are then used to perform the classification. By exploiting the intrinsic properties of quantum computation, such as superposition and entanglement, the team aims to identify if a Jet contains a hadron formed by a $b$ or $\bar{b}$ quark at the moment of production. The approach could lead to new insights and enhance the classification performance in particle physics experiments. Two datasets have been used in this research: the complete dataset and the muon dataset, both of which belong to \ac{CERN}. In the muon dataset analysis, 60,000 Jets are used for training and 40,000 Jets are used for testing. The muon dataset is a subset of the complete dataset, and it is used to evaluate the dependence of the quantum algorithms' performance on the number of training events and the circuit complexity.  The researchers compare the performance of their \ac{QML} approach with that of \ac{DNN}, \ac{LSTM}, and \ac{LSTM} with convolutional layer models. They show that the results for tagging power as a function of the Jet \ac{$p_T$} and $\eta$ are comparable within the  \ac{MSE} error, and therefore, they consider only the \ac{DNN} model for comparison with \ac{QML} algorithms.

\begin{table*}[h!]
\centering
\tiny
\caption{A summary of available ML and \ac{DL} architectures for Jet \ac{HEP} classification, including columns for biases, generalizability, and recommended use cases. Bias levels range from moderate (limited datasets) to high (overfitting, dataset reliance), while generalizability is categorized as high (broad applicability), moderate (adequate performance with some limitations), and low (poor performance or untested on other tasks).}
\label{tab:archCmp}
\begin{tabular}{m{0.5cm}m{0.5cm}m{1.3cm}m{0.6cm}m{0.6cm}m{0.6cm}m{0.6cm}m{0.6cm}m{0.6cm}m{0.7cm}m{0.5cm}m{0.5cm}m{0.7cm} m{3.2cm}}
\hline
Ref. & Year & Model & {IN} & Acc. TT & AUC TT & Acc. QG & AUC QG & Acc. Other & AUC Other & Link & {Biases} & {General.} & {Recommended scenarios} \\
\hline
\cite{pearkes2017jet} & 2017 & TopoDNN & {Image} & 0.916 & 0.972 & -- & -- & -- & -- & No & {M} & {L} & {Top quark identification} \\
\cite{macaluso2018pulling} & 2018 & CNN tagger & {Image} & -- & -- & -- & -- & 0.87 (DTJ) & 0.943 (DTJ) & No & {H} & {H} & {Jet substructure} \\
\cite{komiske2019energy} & 2019 & PFN-ID & {PC} & 0.932 & 0.981 & 0.900 & -- & -- & -- & No & {L} & {L} & {Energy flow studies} \\
\cite{bogatskiy2020lorentz} & 2020 & LGN & {PC} & 0.929 & 0.964 & 0.803 & 0.832 & -- & -- & Yes\footnote{https://github.com/fizisist/LorentzGroupNetwork} & {L} & {M} & {Lorentz invariance studies} \\
\cite{qu2020jet} & 2020 & ParticleNet & {PC} & 0.940 & 0.985 & 0.840 & 0.911 & -- & -- & No & {M} & {H} & {Point cloud analysis} \\
\cite{satorras2021n} & 2021 & EGNN & {PC} & 0.922 & 0.976 & 0.803 & 0.880 & -- & -- & Yes\footnote{https://github.com/vgsatorras/egnn} & {L} & {M} & {Graph neural networks} \\
\cite{mikuni2021point} & {2021} & {PCT} & {PC} & {0.940} & {0.985} & {0.841} & {0.914} & -- & -- & {No} & {L} & {H} & {Point cloud processing} \\
\cite{gong2022efficient} & 2022 & LorentzNet & {PC} & 0.942 & 0.986 & 0.844 & 0.915 & -- & -- & No & {L} & {M} & {Lorentz group studies} \\
\cite{qu2022particle} & 2022 & PartT & {PC} & 0.944 & 0.987 & 0.852 & 0.923 & -- & -- & Yes\footnote{https://github.com/jet-universe/particle\_transformer} & {L} & {H} & {Analysis of long-range feature dependencies in particles}\\
\cite{bogatskiy2022pelican} & {2022} & {PELICAN} & {PC} & {0.942} & {0.986} & -- & -- & -- & -- & Yes\footnote{https://github.com/abogatskiy/PELICAN} & {M} & {L} & {Particle cloud matching} \\
\cite{ruhe2024clifford} & {2024} & {CGENNs} & {PC} & {0.942} & {0.986} & -- & -- & -- & -- & Yes\footnote{https://github.com/DavidRuhe/clifford-group-equivariant-neural-networks} & {L} & {H} & {Clifford group analysis} \\
\cite{spinner2024lorentz} & {2024} & {L-GATr} & {PC} & {0.942} & {0.987} & -- & -- & -- & -- & Yes\footnote{https://github.com/Qualcomm-AI-research/geometric-algebra-transformer} & {M} & {H} & {Geometric algebra studies} \\
\cite{wu2025jet} & {2024} & {MIParT-L} & {PC} & {0.944} & {0.987} & {0.853} & {0.923} & -- & -- & Yes\footnote{https://github.com/jet-universe/particle\_transformer} & {L} & {H} & {Analysis of long-range feature dependencies in particles} \\
\hline
\end{tabular}
\begin{flushleft}
Abbreviation: Input nature (IN); Point cloud (PC);  Top tagging (TT);  Quark-gluon (QG); DeepTop Jets (DTJ); CMS Jets (CJ); Moderate (M); High (H); Low (L).
\end{flushleft}
\end{table*}

\subsection{{MLP and DNN-based methods}}

\begin{figure*}
    \centering
    \includegraphics{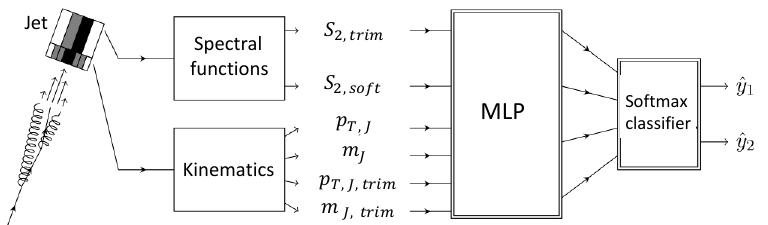}
    \caption{An example of classifier utilizing \ac{MLP} trained using kinematic and spectrum variables  for Jet classification \cite{chakraborty2019interpretable}. $S_{2,trim}$ and $S_{2,soft}$ correspond to hard and soft
substructure information. }
    \label{fig:mlp}
\end{figure*}

\Ac{MLP} is an artificial neural network composed of multiple layers of nodes, including an input layer, one or more hidden layers, and an output layer. Each node in one layer is connected to every node in the subsequent layer. \ac{MLP} can handle complex nonlinear relationships between input and output data, making them suitable for various tasks. \acp{MLP} are versatile, scalable, and can be trained using back-propagation, enabling them to learn from large datasets effectively and generalize well to unseen data. Kinematic parameters describe the motion of particles, including velocity, momentum ($p_{T,J}$) and trimmed Jet momentum ($p_{T,J,trim}$), energy, Jet mass $m_J$ and Jet mass trimmed $m_{J,trim}$ , and angles of emission, commonly used in physics and engineering analyses. Chakraborty et al. in \cite{chakraborty2019interpretable} employed both kinematics and spectral function,  which typically refers to a function that describes the distribution of energy or momentum states of particles in a particular physical system, to feed \ac{MLP} classifier as described in Figure \ref{fig:mlp}. The authors aim is to trim/discard Jet that are unlikely to have originated from the process of interest (effects of background noise). This selective removal helps to improve the accuracy of measurements and analyses by focusing only on the most relevant particles within a jet.

\begin{figure}
    \centering
    \includegraphics[scale=0.6]{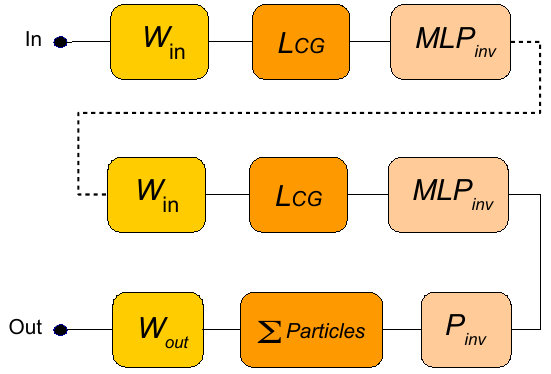}
    \caption{The architecture of LGN model suggested in \cite{bogatskiy2020lorentz}.}
    \label{fig:lng}
\end{figure}

The paper \cite{bogatskiy2020lorentz} introduces the \ac{LGN} neural network model designed for particle physics identification. This model is characterized by its full equivariance to transformations under the Lorentz group, which represents a crucial symmetry of space-time in physics and allows for equivariant nonlinearity. The \ac{LGN} architecture has been successfully applied to a classification task in particle physics called top tagging, whose objective is to distinguish top quark "Jets" from a backdrop of lighter quarks. The \ac{LGN} model consists of several layers, including the linear input layer ($W_{in}$), iterated \ac{CG} layers (\textit{$L_{CG}$}), and the perceptrons $\mathrm{MLP}_{inv}$ layer. This design reduces the number of learnable parameters and provides a deeper understanding of the physical interpretation of the results (Figure \ref{fig:lng}). The initial linear layer processes the 4-momenta of $N_{obj}$ particles originating from a collision event, and it can also handle associated scalar quantities like label, charge, spin, and more. The iterated \textit{$L_{CG}$} layers are defined by a \ac{CG} decomposition of the tensor product of representations of the Lorentz group, which allows for equivariant non-linearity. The \ac{CG} layers are alternated with perceptrons $\mathrm{MLP}_{inv}$ layer, which act only on Lorentz invariants. At the end of each \ac{CG} layer, a \ac{MLP} is applied to the isotypic component of the tensor product. The \ac{MLP} accepts $N^{(p)}_{ch}$ scalar inputs and generates an equivalent number of outputs, with its parameters uniformly applied across all $N_{obj}$ nodes within the \ac{CG} layer. The output layer computes the arithmetic sum of the activations from $N_{obj}$ and extracts the invariant isotypic aspect of this sum. It subsequently employs a final fully connected linear layer, denoted as $W_{out}$, on the $N^{(NCG)}_{ch}$ scalars, generating two scalar weights for binary classification. 
In the \ac{LGN} model's output layer, $P_{inv}$ conducts the projection onto invariants, combines contributions from particles to ensure permutation invariance and subsequently applies a linear transformation. $P_{inv}$ operates independently on each individual particle but maintains consistent parameter values across all particles. The \ac{LGN} model has demonstrated competitive performance while using between 10-1000 times fewer parameters than other \ac{SOTA} models.

{ The \acp{DNN} are a type of artificial neural network that are composed of multiple layers of nodes (i.e., \ac{MLP} with multiple hidden layers)}, with each node connected to every node in the previous and next layers. They are particularly well-suited for processing high-dimensional data, such as images or collections of features, and can learn complex non-linear relationships between inputs and outputs. In the context of  \ac{HEP}, the \ac{DNN} was used in  to classify hadronic Jets based on their input features.  \acp{DNN} typically require a fixed-size input, which can be a limitation when working with variable-length inputs such as particle lists. 

\begin{figure}[ht!]
    \centering
    \includegraphics[scale=0.75]{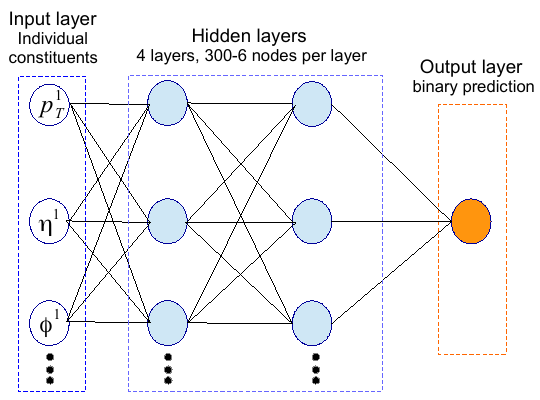}
    \caption{The architecture of the TopoDNN model,  consists of 4 layers with 300, 102, 12, and 6 nodes, respectively \cite{pearkes2017jet}.}
    \label{fig:topodnn}
\end{figure}

In \cite{baldi2016jet}, \acp{DNN} are used in \ac{HEP} to classify Jets produced in particle collisions. \acp{DNN} can automatically extract features from Jet tagging, allowing for more accurate classification than traditional methods that rely on expert-designed features. Parton shower in \ac{HEP} refers to the process where high-energy particles, such as quarks and gluons, emit further particles as they evolve, simulating the fragmentation and radiation patterns observed in particle collisions within particle accelerators, which is crucial for understanding particle interactions. Barnard et al., \cite{barnard2017parton} advocate for \acp{DNN} as hadronic resonance taggers, trained on Jet tagging generated from different generators. The \ac{DNN} showed improved performance on test events generated by the default PYTHIA shower instead of using HERWIG and SHERPA generators, suggesting acquisition of PYTHIA-specific features. However, they noticed that biases may arise from generator approximations. They examine parton shower variations' impact on tagger performance using \ac{LHC} data. Results show up to 50\% differences in background rejection. They introduced the "zooming" method, enhancing performance between 10-20\% across Jet transverse momenta. The TopoDNN model proposed in \cite{pearkes2017jet} is a \ac{DNN}-based architecture (Figure \ref{fig:topodnn}). The network's input layer is designed to process vectors containing the Jet constituents' \ac{$p_T$}, $\eta$, and $\phi$ values. Manual tuning of the network's architecture involved adjusting the depth and node count per layer, within a range of 4-6 layers and 40-1000 nodes per layer, respectively. \ac{ReLU} activation function was implemented in the hidden layers, whereas a sigmoid function was applied to the output node. The training process utilized the Adam optimizer, with training sessions capped at a maximum of 40 epochs. An early stopping mechanism was employed, utilizing a patience parameter set to 5 epochs based on the validation set loss. The final architecture selected features 4 hidden layers, comprising 300, 102, 12, and 6 nodes in each layer respectively.  TopoDNN achieved a significant background rejection of 45 at a 50\% efficiency operating point for reconstruction-level Jets, yielding to correctly identify top quark Jets with a high level of accuracy while rejecting a large portion of background events.

\color{black}

\begin{figure*}[ht!]
    \centering
    \includegraphics[scale=1.2]{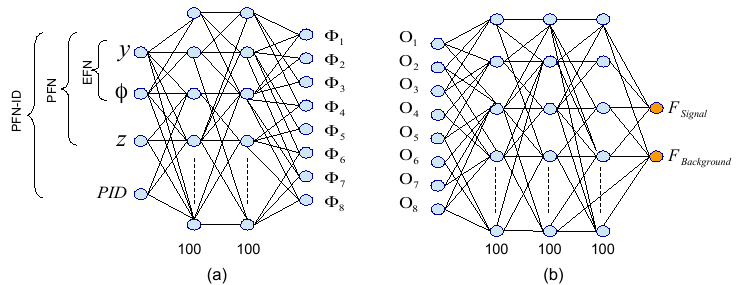}
    \caption{The architecture of PFN-ID model suggested in \cite{komiske2019energy}. (a) Per-particle mapping $\Phi$. (b) The binary output signal or background can be identified.}
    \label{fig:pfnid}
\end{figure*}

The researchers in \cite{lee2018multi} discusses the application of \acp{DNN} to a wide range of physics problems, particularly in \ac{HEP}. Specifically, \acp{DNN} have been successfully applied to tasks such as Jet tagging and event classification. The authors explore the use of a simple but effective preprocessing step that transforms observational quantities into a binary number with a fixed number of digits, representing the quantity or magnitude in different scales. This approach has been shown to significantly improve the performance of \acp{DNN} for specific tasks without complicating feature engineering, particularly in b-Jet tagging using daughter particles' momenta and vertex information. However, the authors in \cite{komiske2019energy} used \acp{DNN} to process collections of ordered inputs, which can be thought of as a fixed-size representation of variable-length inputs. This allows the \ac{DNN} to learn features sensitive to particle ordering, which can be important for discriminating between different types of Jets. \Ac{PFN-ID} model \cite{komiske2019energy} is another proposed type of \ac{DL} architecture that takes particles as input and processes them in a way that is dependent on the order the particles were fed into the network. The \ac{PFN-ID} architecture is based on the Deep Sets framework and includes full particle ID information (Figure \ref{fig:pfnid}). The Deep Sets framework is a \ac{ML} approach that allows for learning directly from sets of features or "point clouds". The following are the main steps of the framework: (i) Map each element of the set to a latent space using a shared function. (ii) Aggregate the latent representations of the elements using a symmetric function. (iii) Map the aggregated latent representation back to the output space using a shared function.  An additive latent space can be used to express a general symmetric function, as provided by the framework. Within the scope of particle-level collider observables, the process involves mapping each particle to a latent representation, which is subsequently collected. Subsequently, the observables are expressed as functions on this latent space. This decomposition includes a diverse range of current collider observables and representations at the event and Jet levels, including as image-based and moment-based techniques. The \ac{PFN-ID} improves the classification performance of the \ac{PFN} model for discriminating quark and gluon Jets. Results show that \ac{PFN-ID} slightly outperforms \ac{RNN}-ID, whereas the \ac{PFN} and \ac{RNN} are comparable.  

The authors in 
\cite{lu2021sparse} introduce a novel \ac{DNN} model, called  sparse autoregressive model (SARM), that learns data sparsity explicitly, yielding stable and interpretable results compared to \acp{GAN}. In two case studies, the first, referred to as $D+D$, employs a discrete mixture model by discretizing pixel values using predetermined grid points, while the second, $D+C$, utilizes a discrete mixture model constructed with a truncated logistic distribution for pixel modeling. In two case studies, SARM outperforms GANs by 24-52\% and 66-68\% on images with high sparsity.

In the study conducted by \cite{fedkevych2023identification}, the identification of $b$ Jets was investigated utilizing \ac{QCD} inspired observables. The process entails the utilization of Jet substructure observables, including one-dimensional Jet angularities and the two-dimensional \ac{PLP}. 
\acp{DNN} are employed to identify $b$ Jets using these \ac{QCD}-inspired observables. The \acp{DNN} are trained on a set of input features, which include Jet angularities and the \ac{PLP}, in order to efficiently distinguish $b$ Jets from light ones. The performance of the \acp{DNN} is evaluated by comparing their results with those of conventional track-based taggers, such as JetFitter, IP3D, and DL1 taggers. In this study, the results indicate that the \ac{DNN} discriminants exhibit better performance than the IP3D tagger.

\color{black}

\subsection{CNN-based methods}

\begin{figure*}
    \centering
    \includegraphics[scale=0.9]{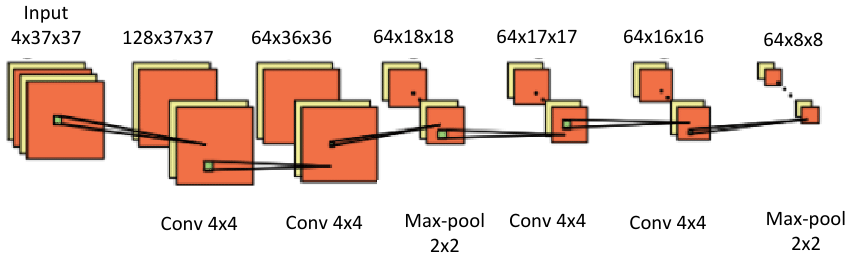}
    \caption{The architecture of CNN tagger model suggested in \cite{macaluso2018pulling}.}
    \label{fig:cnnTagger}
\end{figure*}

\Acp{CNN} have revolutionized Jet image classification and prediction in particle physics.  \acp{CNN} excel in image recognition by leveraging convolutional layers, weight sharing, and pooling to capture hierarchical features, enabling effective pattern recognition and classification \cite{kheddar2023deep,habchi2023ai}. This enables precise particle identification using Jet images, improved event classification, and deeper insights into \ac{HEP} experiments, advancing researchers' understanding of fundamental particles and interactions. For example, the authors in \cite{komiske2017deep} investigate the capability of \acp{CNN} in discriminating quark and gluon Jets, comparing their performance to traditionally designed physics observables. In the realm of Jet image classification, researchers proposed combining \ac{CNN} with various other \ac{DL} techniques. For instance, Farrell's paper \cite{farrell2017hep}, hybrid \acp{DL} revolutionize particle tracking. \acp{LSTM} excel in sequential data analysis, replacing Kalman filtering for hit assignment, while \acp{CNN} construct valuable detector data representations. Their fusion unveils a potent end-to-end model, with GPU training addressing traditional tracking algorithm scaling challenges. 
The \ac{CNN} tagger architecture proposed in the paper  \cite{macaluso2018pulling}   consists of a \ac{CNN} with four identical convolutional layers, each with 8 feature maps and a $4\times4$ kernel. These layers are separated in half by one $2\times2$ max-pooling layer. The CNN also includes three fully connected layers of 64 neurons each and an output layer of two softmax neurons. Zero-padding is included before each convolutional layer to prevent spurious boundary effects. The architecture ends with a flatten layer and three fully connected layers with sizes 64, 256, 256, and 2, respectively (Figure \ref{fig:cnnTagger}). 
The \ac{CNN} is trained on a total of 150k+150k top and \ac{QCD} Jet images, by minimizing a \ac{MSE} loss function using the stochastic gradient descent algorithm in mini-batches of 1000 Jet images and a learning rate of 0.003.  

\begin{figure*}[ht!]
    \centering
    \includegraphics[scale=0.85]{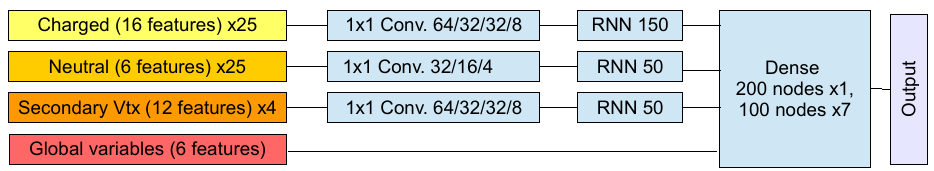}
    \caption{The architecture of DeepJet model suggested in \cite{bols2020jet}.}
    \label{fig:deepjet}
\end{figure*}

Oliveira et al. \cite{de2016jet} applied a \ac{CNN} directly to Jet tagging, showcasing its effectiveness as a powerful tool for identifying boosted hadronically decaying $W$ bosons amid \ac{QCD} multi-Jet processes. Similarly, in order to discriminate Quark-Gluon Jet, Lee et al. in their research \cite{lee2019quark}, employed various pretrained \ac{CNN} models, including VGG, ResNet, Inception-ResNet, DenseNet, Xception, Vanilla ConvNet, and Inception-ResNet, to classify Jet images for distinguishing quark and gluon hadron Jets. The study reveals that DenseNet outperforming larger, higher-structured networks. Despite marginal improvements over a traditional \ac{BDT}  classifier, stability in training can be enhanced using the RMSProp optimizer, an adaptive learning rate optimization algorithm. Similarly, significant progress resulted from integrating 1D \ac{CNN} and \ac{LSTM}, resulting in DeepJet NN model \cite{novak2020sissa} for Jet identification. The architecture extract abstract features from three input collections—secondary vertices, charged particles (tracks), and neutral particles. The final Jet flavor probabilities are determined by combining outputs with global Jet features in dense layers. This architecture was also applied to heavy flavour classification, with the model further adapted for quark-gluon tagging tasks \cite{bols2020jet}. 
In \cite{bols2020jet}, the model architecture consists of several components: (i) Automatic feature extraction is conducted for each constituent through convolutional branches that include $1\times 1$ convolutional layers. Distinct convolutional branches are allocated for vertices, charged particle flow candidates, and neutral particle flow candidates, (ii) the output of the convolutional branches is used to construct a graph representation of the Jet, where each constituent is represented as a node in the graph. The edges between the nodes are determined by a distance metric that takes into account the kinematic properties of the constituents, (iii) the graph representation of the Jet is then processed by several graph convolutional layers, which are designed to capture the correlations between the constituents. The graph convolutional layers use a learnable filter that is applied to the graph representation of the Jet, and (iv) the output of the graph convolutional layers is then fed into several dense layers, which are designed to perform the final classification task. The dense layers use a combination of fully connected and batch normalization layers. In the context of the DeepJet model, the \ac{RNN} layer is an important component of the DeepJet model (Figure \ref{fig:deepjet}), as it allows the model to capture the sequential information in the charged particle tracks and to use this information to improve the classification performance. The DeepJet model has been shown to achieve \ac{SOTA} performance in Jet flavour classification and quark/gluon discrimination tasks. The model was tested using \ac{CMS} simulation and was found to outperform previous classifiers, including the IP3D algorithm. The DeepJet model underwent a comparative analysis against a binary quark/gluon classifier from the \ac{CMS} reconstruction framework. An improvement in performance was noted with the use of the DeepJet model on a dataset comprised exclusively of light quark and gluon Jets. Moreover, the DeepJet model was found to be more robust to variations in the Jet constituents and kinematics, which makes it more suitable for use in real-world scenarios. In terms of DeepJet's Performance, using the function of reconstructed vertices, b-Jet efficiency can reach 92\%, and when the function of Jet \ac{$p_T$}, b-Jet efficiency is around 95\%. 

Du et al. in their paper \cite{du2012deep} addressed challenges in assessing Jet distribution modification in a hot \ac{QCD} medium during heavy-ion collisions. It utilizes a \ac{CNN} trained on a hybrid strong/weak coupling model, achieving good performance and emphasizing result interpretability. The study reveals discriminating power in the angular distribution of soft particles and explores the potential of \ac{DL} for tomographic studies of Jet quenching.

\begin{figure*}[ht!]
    \centering
    \includegraphics{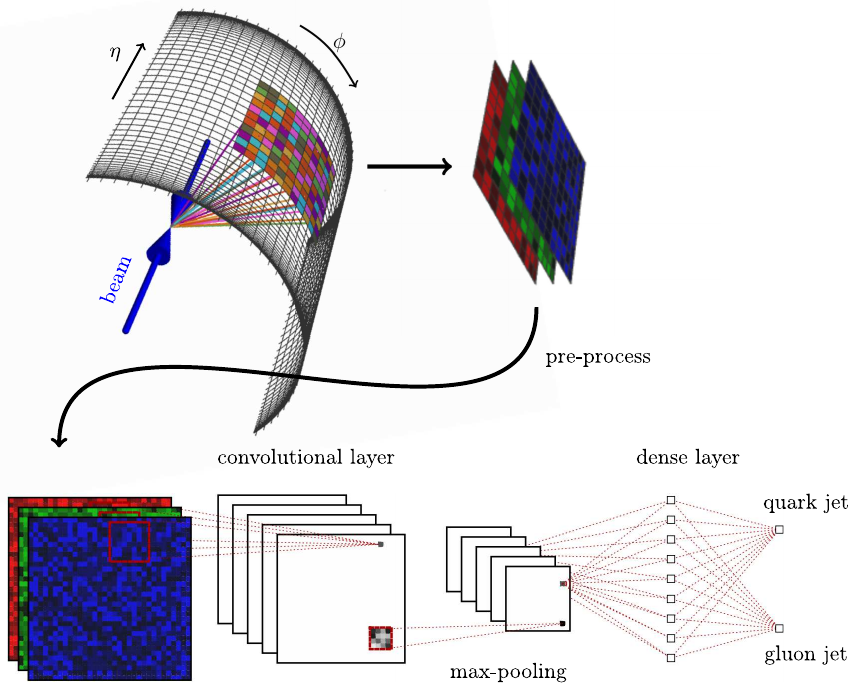}
    \caption{Example of CNN architecture with input Jet image, three convolutional layers, dense layer, and output layer are involved. In this context, \textbf{red} represents the transverse momenta of charged particles, \textbf{green} corresponds to the \ac{$p_T$} of neutral particles, and \textbf{blue} signifies the charged particle multiplicity \cite{komiske2017deep}.}
    \label{fig:jetdiscrimination}
\end{figure*}

The study \cite{du2021classification} demonstrates \ac{CNN}'s efficacy in predicting energy loss for quark and gluon Jets, yielding comparable results. It highlights distinctions post-quenching and employs \ac{DL} for classification, emphasizing energy loss's impact on classification difficulty.    In Figure \ref{fig:jetdiscrimination}, a CNN architecture is presented specifically designed for identifying quark and gluon Jets.   The researchers \cite{kim2023multi} employed \ac{CNN} to analyze \ac{LHC} proton-proton collision simulation data. Their \ac{CNN} model, utilizing detector responses as images, distinguishes r-parity violating super-symmetry (RPV SUSY) signal events from \ac{QCD} multi-Jet background events. Achieving 1.85 times efficiency and 1.2 times expected significance over traditional methods. the authors showcased the model's scalability on HPC resources, reaching 1024 nodes.

\subsection{Adversarial training-based methods}

\Acp{GAN} in image processing enhance creativity and realism by generating new images through a dynamic interplay. The generator creates images, while the discriminator evaluates and refines them, enabling tasks like image-to-image translation, style transfer, and data augmentation with unparalleled versatility \cite{rehm2021reduced,habchi2023ai}. \Acp{GAN} are powerful tools for Jet image classification in particle physics.  They create realistic Jet images, enabling robust testing of classification algorithms. \acp{GAN} enhance the accuracy of identifying particles and contribute to breakthroughs in \ac{HEP} research. However, the authors in 
\cite{stein2022improving} employed another technique for adversarial training for physics object identification and decreased the effect of simulation-specific artifacts. They systematically distorted inputs that have been generated with \ac{FGSM}
adversarial attack technique, this latter altering model predictions using gradient information. The method showed how model performance and robustness are related. They explored the trade-off between performance on unperturbed and on distorted test samples, investigating ROC curves and \ac{AUC} scores for the used discriminators. Similarly, in \cite{stein2023improving},  the paper investigates the loss manifold of a Jet tagging algorithm concerning input features on nominal and adversarial samples. Discrepancies in flatness reveal differences in robustness and generalization. The study suggests refined training approaches through macro-scale loss manifold exploration for two features and devising attacks that maintain the gradient's directionality. This leverages acquired insights for enhanced object identification in particle physics.

%\cite{lawrance2022generation}

%\cite{cheng2023predicting}

\subsection{RNN-based methods}

Various types of \acp{RNN} such as \acp{BRNN}, \ac{LSTM}, and \acp{GRU} differ in architecture at the cell level within the RNN layer. \acp{BRNN} propagate information in both forward and backward directions, influencing predictions by surrounding words. \ac{LSTM} tackles vanishing gradients with inner cells containing input, output, and forget gates, regulating information flow. \acp{GRU}-based networks address short-term memory issues with reset and update gates controlling information utilization akin to \ac{LSTM} gates \cite{auricchio2023vbf,kheddar2023deep}.
\Acp{RecNN}, are designed to operate on hierarchical or tree-structured data, where the relationships between elements are defined by a recursive structure. Instead of processing sequences sequentially with temporal dependencies, like \acp{RNN}, \Acp{RecNN} recursively apply the same neural network operation to combine representations of child nodes to produce a representation of their parent node, traversing the hierarchical structure. In light of this, the authors in \cite{cheng2018recursive} investigate \Acp{RecNN} for quark/gluon discrimination. Results indicate \Acp{RecNN} outperform baseline, boosted decision tree,  in gluon rejection rate by a few percent. Even with minimal input features such as \( p_T, \eta, \phi \), \Acp{RecNN} yield promising results, suggesting tree structure contains essential discrimination information. Additionally, rough up or down quark Jet discrimination is explored. In \cite{auricchio2023vbf}, a neural network was created specifically for Jet binary classifying .  The network comprises two hidden layers employing recurrent cells, with a structure consisting of 25 \Ac{LSTM} cells and utilizing a tanh activation function at its core.

%\subsection{MLP-based methods}

\begin{table*}[ht!]
\scriptsize
\caption{Summary of the performance of certain \ac{ML} and  \ac{DL} frameworks proposed for \ac{HEP}. Only the best performance is reported in the case of multiple tests.} \label{tab:FL}

\renewcommand{\arraystretch}{1.6} 
\begin{tabular}{m{0.5cm}m{1cm}m{1.5cm}m{4.5cm}m{1.5cm}m{6cm}m{0.5cm}}
  \toprule
   Ref. & DLM & Dataset & Description & BP (\%)  & Limitations & PLA  \\ \hline
   \toprule
   \cite{kim2023multi} &  CNN &  QCD multi-Jet  &  Classification of multi-Jet events using \ac{CNN} at high energies of 13 TeV  &  AUC= 99.03  &  The proposed \ac{CNN} model needs validation with additional datasets to ensure its generalizability.  & No  \\

    \cite{munoz2022boost} &  SVM &  Simulated  &  \Ac{BIP} features invariant under boosts for improved Jet tagging  &  Acc= 92.7  &  Performance could be enhanced through comprehensive hyperparameter tuning.  & Yes\tablefootnote{\url{https://zenodo.org/records/7271316}}  \\
   
    \cite{pearkes2017jet} & DNN & Simulated & Sequence of Jet components arranged in a specific order for training inputs. & Acc= 50 & Could be enhanced by employing the LSTM method to efficiently classify Jet from background. & No\\

    \cite{baldi2016jet}  & DNN & Simulated\tablefootnote{\url{https://www.igb.uci.edu/~pfbaldi/physics/}} & \acp{DNN} for categorizing Jet substructure in \ac{HEP} & AUC= 95.3 & The accuracy of the \ac{DNN} models is limited by the accuracy of the simulation models used to generate the training data.& No\\
    
    \cite{pezoa2023explainability} &  DNN &  Higgs   &  Clarifying \ac{HEP} event classification with \ac{SHAP}  &  Acc= 66  & \ac{SHAP} may not comprehensively capture feature interactions or explain model behavior in all cases. It could demand substantial computational resources for large datasets or intricate models. & Yes\tablefootnote{\url{https://github.com/rpezoa/hep_shap/}} \\
    
    \cite{fedkevych2023identification} &  DNN &  ATLAS   &  Detection of b Jets utilizing \ac{QCD}-inspired measurements  &  AUC= 67  &  The \ac{DNN} performed slightly less effectively than the JetFitter algorithm.  & No  \\

    \cite{lu2021sparse} & DNN & Pythia \newline Jet images & Creating images with low pixel density in particle physics for two cased $D+D$ and $D+C$. & AUC= 86.9, \newline AUC= 84.1  & Slower than the non-autoregressive model LAGAN. $D+D$ performed better than $D+C$ for both Pythia and Monte Carlo images.   & Yes\tablefootnote{\url{http://mlphysics.ics.uci.edu/}}\\

    \cite{du2021classification} &  CNN &  Simulated  &  \ac{CNN} for predicting quark and gluon Jets  &  Acc= 75.9  &  The higher the energy loss, the more challenging the task of classifying the Jets becomes.  & No  \\

    \cite{cheng2018recursive} &  RecNN &  Simulated &  Enhance Quark/gluon classification  &  AUC= 86.37  &  Event-level analysis is not performed.  & Yes\tablefootnote{\url{https://github.com/glouppe/recnn}}  \\
        
    \cite{racah2016exploring} &  CNN-AE &  Daya Bay  &  Classification for different event types, including IBD prompt, IBD delay, Muon, Flasher, and other  &  Acc= 99.9 (Muon)  &  SVM and KNN exhibit inferior performance compared to CNN in identifying event types. Moreover, semi-supervised techniques have not been examined. & No  \\
    \cite{chen2022quantum} &  CNN &  Simulated   &  Employing a quantum \ac{CNN} to categorize events in \ac{HEP}.  &  Acc= 97.5  &  Quantum \ac{CNN} showed a lower performance than CNN when it comes to a binary classification of \textbf{Muon} and  \textbf{Electron}. Besides, CNN showed low performance when classifying \textbf{Muon} and  \textbf{Pion} compared to quantum \ac{CNN}. & No  \\
    \cite{caron2017bsm} &  ML &  ATLAS  &  Predict if the LHC trials have dismissed a new physics model  &  Acc= 93.8  &  Enhancing reliability can be achieved by requiring a minimum confidence level for the prediction.  & Yes\tablefootnote{\url{http://susyai.hepforge.org}} \\
    
    \cite{almeida2015playing} &  ANN &  Simulated  &  Identifying boosted top quarks using pattern recognition through an \ac{ANN} in \Ac{HEP} experiments.  &  Eff= 60  &  It has 4\% mis-tag rate. It exclusively utilizes \ac{HCAL} data, though additional data, like sub-Jet b-tags, are crucial for top tagging. & No  \\
     \cite{stoye2018deep} &  DNN &  Real data  &  Enhancing Jet reconstruction at CMS through \ac{DL}  &  FPR= 65  &  The computational costs, wnen employing the proposed model, have not been verified. & No  \\

    \cite{apolinario2021deep} & CNN & Simulated & Detection of Jet quenching effects caused by the presence of the \ac{QGP}. & AUC= 75 & The computational costs, when employing the proposed model, have not been verified. When data normalized, AUC reached only 67\% (when $p_{T,jet} > 30 GeV$). & No \\
            
   \bottomrule
\end{tabular}
\begin{flushleft}
Abbreviations: \ac{DL} model (DLM); Best performance (BP); Project link availability (PLA); auto-encoder (AE).
\end{flushleft}
   
\end{table*}

\subsection{GNN-based methods}

\begin{figure*}
    \centering
    \includegraphics[scale=0.6]{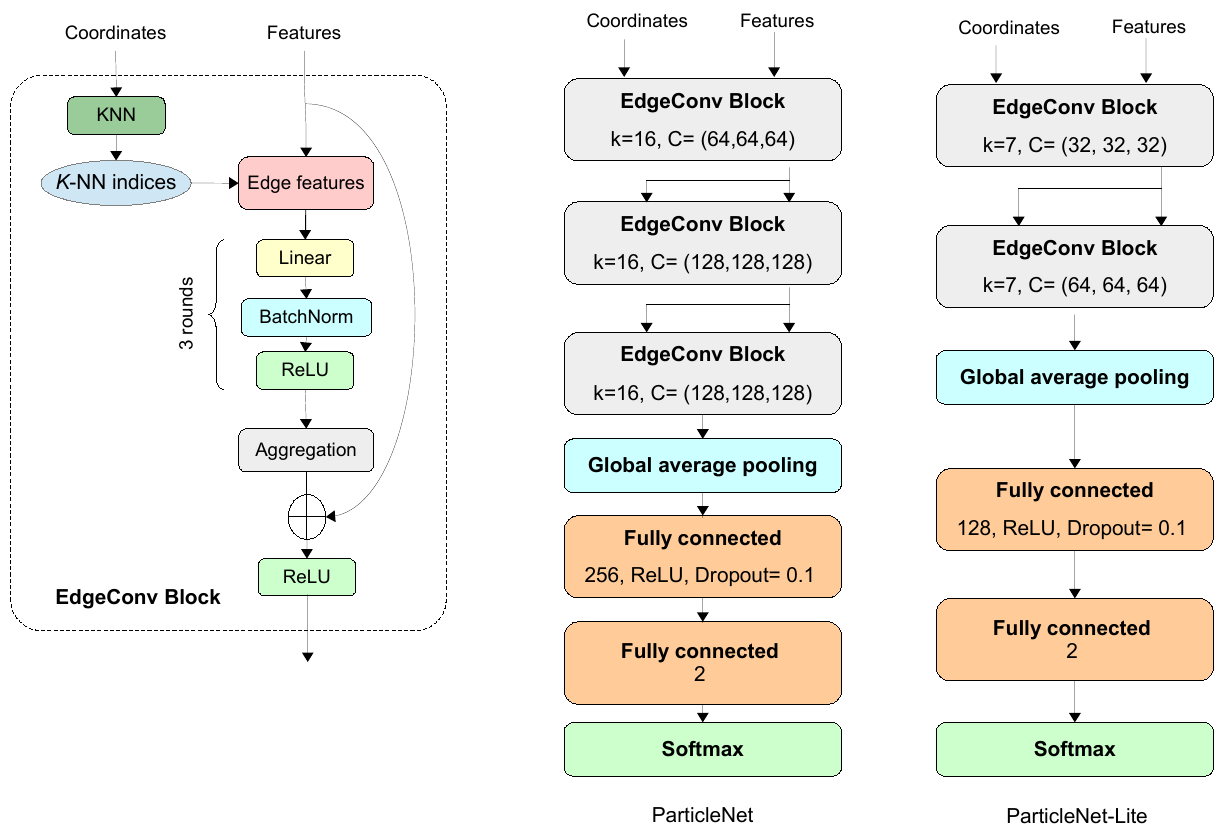}
    \caption{The architecture of ParticleNet model suggested in \cite{qu2020jet}.}
    \label{fig:partNet}
\end{figure*}

\Acp{GNN} are neural networks designed for graph-structured data, learning node and edge representations while capturing complex relationships and dependencies within graphs for tasks such as classification and prediction. In the \ac{HEP} context, the authors in \cite{qu2020jet} proposed the ParticleNet model (Figure \ref{fig:partNet}). The architecture is a customized neural network that operates directly on particle clouds for Jet tagging. It uses dynamic graph \acp{CNN}  to process the unordered set of constituent particles that make up a Jet. The architecture consists of three EdgeConv blocks, each with a different number of channels and nearest neighbors. EdgeConv block starts by representing a point cloud as a graph, whose vertices are the points themselves, and the edges are constructed as connections between each point to its \ac{KNN} points. The EdgeConv block then finds the \ac{KNN}  particles for each particle, using the "coordinates" input of the EdgeConv block to compute the distances. Inputs to the EdgeConv operation, the "edge features," are constructed from the "features" input using the indices of \ac{KNN} particles. The EdgeConv procedure is executed using a three-layer \ac{MLP}. Each layer is structured to include a linear transformation, succeeded by batch normalization, and subsequently a \ac{ReLU} activation. Additionally, a shortcut connection is integrated into every block parallel to the EdgeConv operation, facilitating the direct passage of input features. An EdgeConv block is defined by two key hyper-parameters: the neighbor count $k$ and the channel count $C$, which respectively denote the number of neighbors to consider and the number of units within each layer of linear transformation. The EdgeConv blocks play a crucial role in learning the local features of the particle cloud and aggregating them into a global feature vector for the Jet. Following EdgeConv blocks, global average pooling aggregates particle features, leading to a 256-unit fully connected layer, \ac{ReLU} activation, dropout, and a 2-unit softmax output for binary classification. The ParticleNet architecture achieves \ac{SOTA} performance on two representative Jet tagging benchmarks and is improved significantly over existing methods.

\color{black}

Similarly, \cite{satorras2021n} proposed the \ac{EGNN} model, which is a \ac{GNN} architecture that is translation, rotation, and reflection equivariant (E(n)), and permutation equivariant with respect to an input set of points. It uses a set of filters that are equivariant to the action of the symmetry group, which are constructed using a combination of radial basis functions and Chebyshev polynomials.  
The \ac{EGNN} algorithm possesses the same flexibility as the \ac{GNN} technique, while also maintaining E(n) equivariance similar to the radial field algorithm. Additionally, it eliminates the requirement for computationally intensive procedures, such as spherical harmonics.  The \ac{EGNN} exceeds other equivariant and non-equivariant options while maintaining efficiency in terms of running time. Moreover, the \ac{EGNN} approach demonstrates a 32\% reduction in error compared to the \ac{SOTA} method. 

\begin{figure}
    \centering
    \includegraphics[scale=0.5]{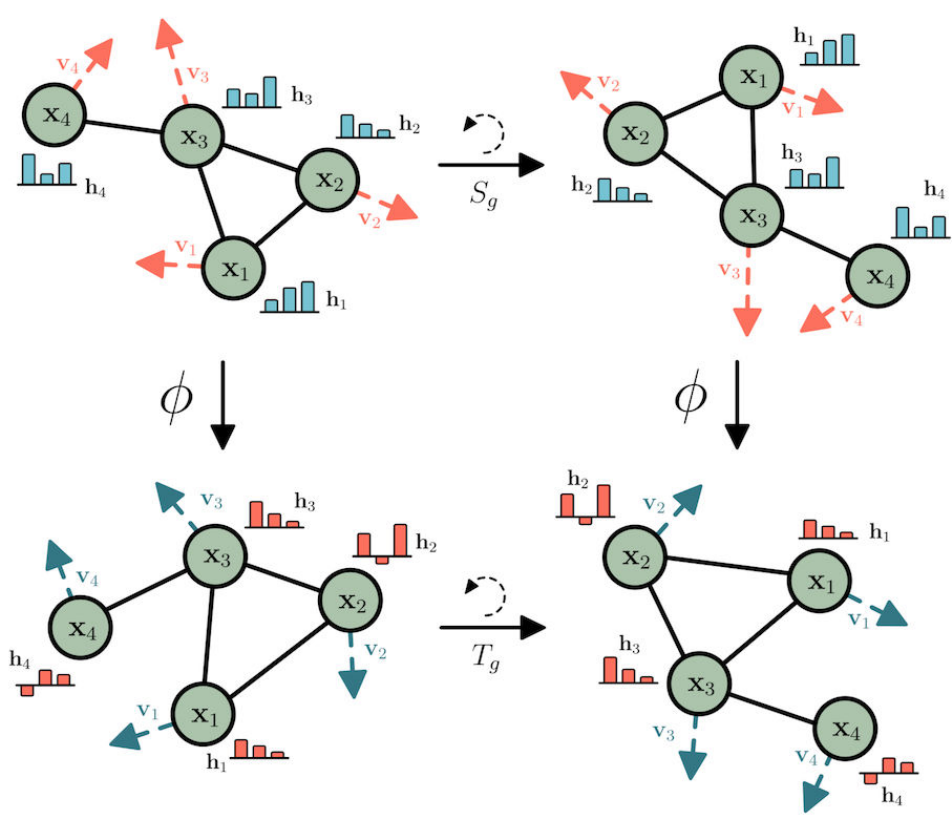}
    \caption{The architecture of EGNN model suggested in \cite{satorras2021n}.}
    \label{fig:egnn}
\end{figure}

Another architecture called the LorentzNet is proposed in \cite{gong2022efficient}, which is based on the \ac{LGEB} block. The structure of \ac{LGEB} consists of several layers, including Minkowski norm and inner product, sum pooling, a \ac{MLP}, and a Clebsch-Gordan tensor product. The input of \ac{LGEB} is a set of 4-momentum vectors, which are transformed by the Minkowski norm and inner product layer to obtain Lorentz-invariant geometric quantities. The sum pooling layer aggregates the geometric quantities to obtain a scalar representation of the input. The \ac{MLP} layer is used to learn a nonlinear mapping from the scalar representation to a new feature space. Finally, the Clebsch-Gordan tensor product layer is used to combine the new feature space with the original input to obtain the output of \ac{LGEB}. It is designed as a Lorentz group-equivariant mapping to preserve the symmetries of the Lorentz group, ensuring the model's equivariance and universality. 

\begin{figure*}[h!]
    \centering
    \includegraphics[scale=1.2]{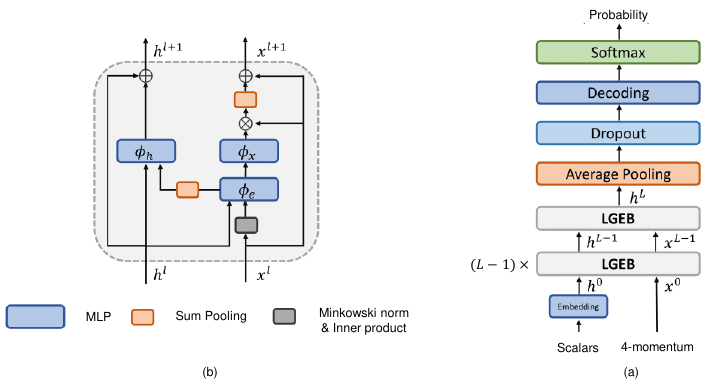}
    \caption{(a) The architecture of LorentzNet model. (b) LGEB block \cite{gong2022efficient}.}
    \label{fig:LorentzNet}
\end{figure*}

The paper \cite{ruhe2024clifford} introduced \acp{CGENN}, a novel \ac{GNN} framework designed to construct \( O(n) \)- and \( E(n) \)-equivariant models using Clifford algebra. \acp{CGENN} leveraged the geometric properties of Clifford algebras, such as the geometric product, to parameterize equivariant neural network layers. These layers operated on multivectors—structures encompassing scalars, vectors, and higher-dimensional geometric features—enabling symmetry-aware computations. Input point cloud included scalars (e.g., mass) and vectors (e.g., positions), embedded into multivector subspaces. \acp{CGENN} achieved \ac{SOTA} performance across domains, including 3D \( n \)-body simulations and 4D Lorentz-equivariant tasks, and Jet tagging in \ac{HEP}, outperforming models like LorentzNet and \ac{EGNN}. However, their computational costs, due to complex geometric products, remained a challenge for scalability and real-time applications.

\color{black}

\subsection{{Transformer-based methods}}
Transformers are \ac{AI} models using self-attention mechanisms to process sequential data, excelling in natural language processing \cite{djeffal2023automatic}, computer vision \cite{habchi2024machine}, and time-series tasks by capturing long-range dependencies and contextual relationships efficiently. Researchers in \ac{HEP} have investigated transformers for the Jet tagging task.  For example, {\cite{mikuni2021point} introduced a modified \ac{PCT} for jet-tagging tasks in collider physics. The \ac{PCT} leveraged self-attention layers and EdgeConv blocks to handle the unordered nature of particle data, ensuring permutation invariance. Jets were represented as point clouds with up to 100 particles, described by kinematic features such as momentum and particle types. The suggested \ac{PCT}  achieved \ac{SOTA} performance, with an high \ac{AUC}  for both  top tagging and quark-gluon classification, showing up to a 20\% improvement in background rejection over models like ParticleNet. Despite its superior performance, the computational cost was significant, with 266M FLOPs, making real-time applications challenging.}

In addition, the work in \cite{qu2022particle} proposed PartT, which is a new Transformer-based architecture for Jet tagging. Its main task is to identify the origin of a Jet of particles produced in \ac{HEP} experiments. ParT makes use of two sets of inputs: (i) the particle input, which includes a list of features for every particle and forms an array, and (ii) the interaction input, which is a matrix of features for every pair of particles. ParT employs a novel \ac{P-MHA} mechanism, which allows the model to attend to pairs of particles and learn their interactions. The \ac{P-MHA} is more effective than standard plain multi-head attention. This assertion is substantiated when the pre-trained ParT models are fine-tuned on two widely adopted Jet tagging benchmarks, the quark-gluon tagging dataset and the binary classification dataset for identifying boosted $W$ bosons decaying to two quarks. The fine-tuning process involves training the ParT models on a smaller labeled dataset specific to each benchmark, which allows the models to learn the specific features and patterns relevant to each task. The fine-tuned ParT models achieve significantly higher tagging performance than the models trained from scratch and outperform the previous \ac{SOTA} models, including ParticleNet and other Transformer-based models.  

\begin{figure*}[h!]
    \centering
    \includegraphics[scale=0.7]{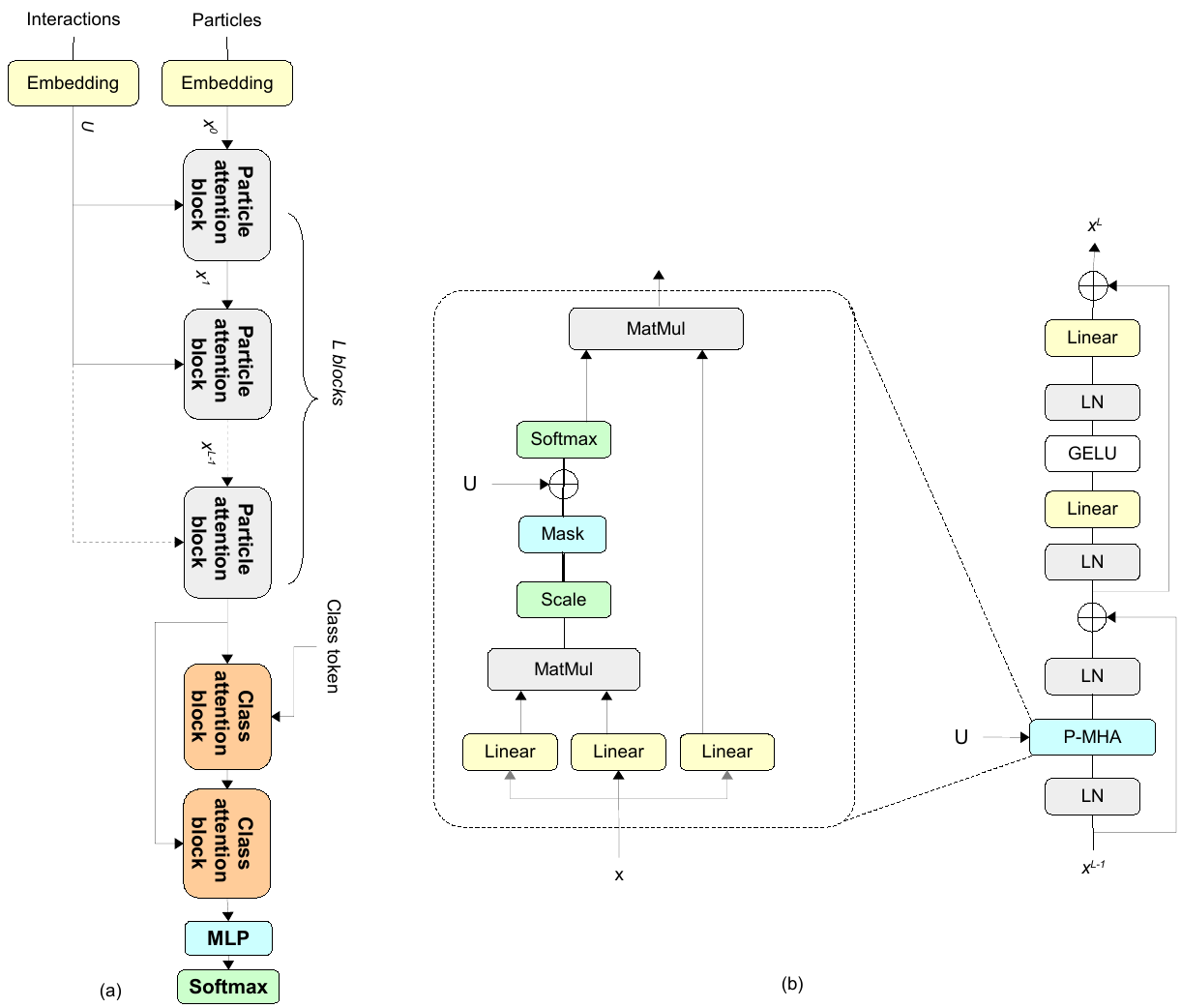}
    \caption{The architecture of PartT model suggested in \cite{qu2022particle}. (a) Particle Transformer; (b) Particle attention block.}
    \label{fig:partT}
\end{figure*}

Moving on, \cite{spinner2024lorentz} introduced the \ac{L-GATr}, a versatile architecture designed for high-energy physics. \ac{L-GATr} combined Lorentz-equivariant geometric algebra with attention mechanisms, enabling robust handling of particle physics data in four-dimensional spacetime. The architecture accommodated variable-length inputs, exploited Lorentz symmetry, and extended to generative modeling via continuous normalizing flows trained with Riemannian flow matching. It used Transformer-based layers with Lorentz-equivariant attention and normalization tailored to Minkowski space, processing particle data parameterized by type and four-momentum vectors. The evaluation employed metrics such as accuracy, \ac{AUC}, background rejection rates, MSE, likelihood, and two-sample tests. \ac{L-GATr} demonstrated competitive or superior performance compared to Lorentz-equivariant graph networks. However, it had computational overhead relative to standard transformers and left its potential for pretraining in \acp{HEP} unexplored. Similarly,  \ac{MIParT} scheme \cite{wu2025jet} introduced the more-interaction Attention (MIA) mechanism to enhance Jet tagging by embedding detailed particle interactions. Based on the Transformer architecture, MIParT-L doubled the dimensions of interaction embeddings for large datasets while reducing model complexity, with 30\% fewer parameters and 53\% lower computational demands than its predecessor, ParT. Tested on top tagging and quark-gluon datasets, MIParT-L achieved nearly identical accuracy and \ac{AUC} to leading models while improving background rejection by 25\% and 3\%, respectively. Fine-tuning on large pre-trained datasets further improved performance by 39\% and 6\%. Despite its efficiency, the interpretability of MIParT-L  remained a challenge, limiting insights into its decision-making process. This trade-off underscored the computational of model  efficiency and robust performance across diverse Jet tagging tasks.

\color{black}

\section{Applications of AI-based Jet {classification}}
\label{sec6}
\begin{figure*}
    \centering
    \includegraphics[scale=0.8]{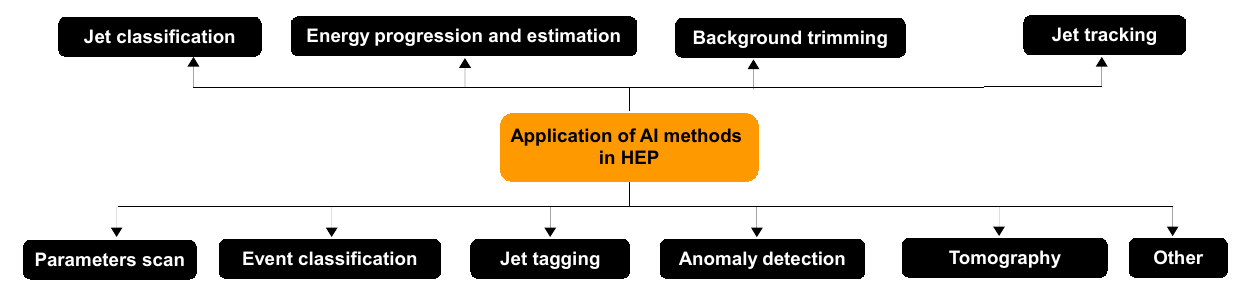}
    \caption{Taxonomy of AI-based HEP applications using Jet images or {PC}.}
    \label{fig:JetApps}
\end{figure*}

Jet images and \ac{PC} processed through \ac{ML} and \ac{DL} techniques hold vast potential across various applications within the \ac{HEP} domain, some of theme are already described in \cite{abdughani2019supervised}. This section presents a comprehensive overview of cutting-edge work in this area, categorized into several key domains: Jet parameter scanning, event classification, Jet tagging, multi-Jet classification, energy estimation, and beyond \cite{kagan2022image}. The taxonomy of AI-based Jet image {and \ac{PC}} applications is visualized in Figure \ref{fig:JetApps}, illustrating their scope and relationships. The section  thoroughly reviews some applications conducted by researchers, while suggesting future directions for those not yet explored.    Additionally, Table \ref{tab:FL} provides a concise summary of performance metrics, limitations, online project availability, and results  obtained across these applications, offering valuable insights into their efficacy and applicability.

\subsection{{Jet} parameters scan}

{A parameter scan in \ac{HEP} involves systematically exploring a wide range of values for the theoretical parameters that define a given model. These parameters often characterize the masses of new particles, coupling strengths, or other fundamental quantities hypothesized in extensions of the \ac{SM}. By examining different combinations of these parameters, researchers aim to identify which sets are compatible with current experimental data or make predictions that can be tested in future experiments. This process helps narrow down the vast theoretical landscape to more plausible scenarios, guiding ongoing investigations and informing the design of new searches \cite{gambit2017gambit}.}

The utilization of \ac{ML} and \ac{DL} models enables the comprehension and estimation of the correlation between the parameter space of new physics models and the experimental physical observables, including signatures characterized by Jets, leptons, and missing transverse energy. This facilitates the efficient constraint of the parameter space of the new physics model \cite{abdughani2019supervised}. Given the sensitivity of the ATLAS experiment to exploring parameters, event counts, and Jet distributions in new physics scenarios, significant computing power is required to deduce the surviving regions of the parameter space of \ac{CMSSM} using Bayesian posterior probability and likelihood function ratio tests. 

To mitigate computational demands, the study \cite{bridges2011coverage} utilizes an \ac{MLP} as a regressor to learn the mapping from \ac{CMSSM} model parameters $\theta$ to weak-scale supersymmetric particle masses $m$. The output of the SoftSusy physical package serves as the target output value of the neural network. Approximately 4000 sample points in the parameter space form the training set. With a given set of \ac{CMSSM} parameters, this \ac{MLP} model rapidly predicts the corresponding supersymmetric particle mass spectrum, which can then be used to forecast observable distributions at the \ac{LHC}, including Jet multiplicities and kinematic features. This approach significantly accelerates the process compared to traditional methods. To identify the parameters of a new physics model, \cite{bornhauser2013determination} trained an \ac{MLP} using 84 physical observables from the 14 TeV \ac{LHC} as inputs, many of which involve Jets and their kinematic properties, with the parameters of a supersymmetric model as the desired outputs. The study revealed that with a collider luminosity of 10 $fb^{-1}$, the \ac{CMSSM} model’s parameters $M_0$ and $M_{1/2}$ could be reliably determined with just a 1\% margin of error. With a collider luminosity of 500 $fb^{-1}$, additional model parameters such as $\tan \beta$ and $A_0$ could also be accurately estimated. In contrast, the conventional approach of minimizing $\chi^2$ yielded comparatively inferior results.

Generating collider event samples at the \ac{LHC} through Monte Carlo simulation can be time-intensive, especially when analyzing detailed Jet structures. While a rapid detector simulation requires only a few minutes, a comprehensive simulation using the GEANT4 framework, as employed by ATLAS and CMS, may necessitate several days. To address this, \cite{buckley2012fast} applied parallel full detector simulations using four parameters—common scalar mass ($m_0$), universal gaugino mass ($m_{1/2}$), the trilinear coupling ($A_0$), and the ratio of vacuum expectation values ($\tan \beta$)—to produce events including Jets and other final-state objects more efficiently. Two \ac{ML} models, the \ac{MLP} and \ac{SVM}, were employed to learn the correlation between the number of signal events and the \ac{CMSSM} parameters. The results showed that predicting the likelihood function, which strongly depends on Jet signatures and other observables, could achieve several percent accuracy with just 2000 training samples. Moving on, the paper \cite{ren2019exploring} proposes a \ac{MLS} framework for efficient exploration of multi-parameter supersymmetric models, surpassing traditional methods like MCMC and MultiNest. Utilizing deep neural networks, the \ac{MLS} incrementally learns parameter space, reducing computational costs while improving target discovery. It integrates \ac{HEP} packages for precise calculations, including tools like GAMBIT and micrOMEGAs, demonstrating efficiency on toy and CMSSM datasets. Achieving up to 80\% sampling efficiency in constrained parameter spaces, \ac{MLS} outperforms MultiNest under 68\% and 95\% confidence levels, offering scalability and adaptability for physics model analysis.

\color{black}

\subsection{Jet classification and tagging}

Despite treating Jets as images or \ac{PC} in the calorimeter and exploiting the benefits of \acp{DNN} in  classification for improved Jet substructure detection, these approaches encounter hurdles. Challenges such as Jet image sparsity and potential precision loss arise from constructing Jet images through pixelation or creating advanced Jet features. In this study \cite{pearkes2017jet}, a sequential method is employed, utilizing an ordered sequence of Jet constituents as inputs for training. Unlike many prior methods, this approach avoids information loss during pixelization or high-level feature computation. The Jet classification technique achieves a considerable background rejection efficiency operating point for reconstructed Jets with transverse momentum ranging from 600 to 2500 GeV. Moreover, it remains unaffected by multiple proton-proton interactions at levels anticipated during Run 2 of the \ac{LHC}. 

Particles generated in a collider with significant center-of-mass energy typically exhibit high velocity. As a result, their decay products tend to align closely, leading to overlapping Jets. It is crucial in collider data analysis to discern whether a Jet originates from a solitary light particle or from the decay of a heavier particle. Traditional approaches rely on manually crafted distribution features based on energy deposition in calorimeter cells. However, due to the intricate nature of the data, \ac{ML} techniques have proven more efficient than human efforts for this task \cite{bhattacherjee2019study}. In \cite{cogan2015jet}, the Jet image concept treats the detector as a camera, capturing Jet energy distribution in calorimeters as a digital image. This enables Jet tagging as a pattern recognition task, utilizing \ac{ML}, like Fisher classification, to differentiate between hadronic W boson decay and Jets from quarks or gluons. Monte Carlo simulation shows superior discrimination compared to traditional methods, offering insights into Jet structure. In \cite{komiske2017deep}, \acp{CNN} improve tagging by treating Jet energy distribution as an image, using channels for features like particle momentum and count. Results show \acp{CNN} can surpass traditional methods, providing reliable insights from collider simulation data despite variations in event generators. However, \acp{CNN} demonstrate a lack of sensitivity to quark/gluon Jets from different generators, akin to conventional Jet measurements. Moving on, in \cite{louppe2019qcd}, Jet tagging is performed using \ac{RNN}, leveraging the similarity between Jet clustering and natural language structure. Final-state particle four-momenta are treated as language words, and Jet clustering as grammatical analysis. \ac{RNN} efficiently processes the tree-like Jet structures, enabling direct use of particle data regardless of count. This method yields higher data utilization efficiency and prediction accuracy than Jet image-based \ac{ML}, extending to event classification. In \cite{cheng2018recursive}, \acp{RNN} distinguish quark and gluon Jets, showing higher gluon suppression. Factors affecting \ac{RNN} performance are explored, with preliminary quark tagging results. Numerous explorations for phenomena beyond the \ac{SM} at the \ac{LHC} depend on top tagging techniques that distinguish between boosted hadronic top quarks and the more prevalent Jets that originate from light quarks and gluons. The \ac{HCAL} essentially captures a "digital image" of each Jet, where the pixel brightness represents the energy deposited in \ac{HCAL} cells. Therefore, top tagging is essentially a matter of recognizing patterns. The work in \cite{almeida2015playing} propose a novel top tagging algorithm based on an \Ac{ANN}, a popular pattern recognition approach. The \ac{ANN} is developed using a substantial dataset of boosted tops along with light quark/gluon Jets and is subsequently evaluated on separate datasets. In Monte Carlo simulations, particularly within the 1100-1200 GeV range, the ANN-based tagger demonstrates outstanding efficacy.

Efficient \ac{HEP} data analysis is imperative with the surge in data from modern particle detectors. However, detectors have limited access to the substructure of Jets, especially those distant from the center-of-mass frame. To address this, the authors \cite{munoz2022boost} integrate \ac{BIP} features with standard classification methods, significantly improving Jet tagging efficiency. Notably, supervised methods like \ac{MLP}, XGBoost, LogReg, \ac{SVM}, and unsupervised approaches like \ac{GMM} and \ac{KNN} achieve exceptional performance with \ac{UMAP} dimensionality reduction technique, surpassing contemporary \ac{DL} systems while reducing training and evaluation times significantly. In \cite{stoye2018deep}, the authors introduce a novel network architecture designed for Jet tagging in experiments conducted at the \ac{LHC}. DeepCSV, currently endorsed by \ac{CMS} and employing a \ac{DNN}, has significantly improved tagging performance, as validated using real collision data. It surpasses other tagging methods, particularly at high transverse momenta, with nearly an order of magnitude reduction in \acp{FPR} using standard threshold definitions.

%\subsection{Multi-Jet classification}

Multi-Jet classification is a  a key task in particle physics aimed at distinguishing between events with varying numbers of Jets. Using \ac{ML} techniques, such as \acp{DNN}, researchers develop classification models to accurately identify these events. Achieving high classification accuracy is crucial for understanding fundamental particle interactions and discovering new physics phenomena in experiments like those conducted at the  \ac{LHC}. The work in \cite{kim2023multi} present an application of scalable \ac{DL} to analyze simulation data from proton-proton collisions at 13 TeV in the \ac{LHC}. The researchers developed a \ac{CNN} model which utilizes detector responses as two-dimensional images reflecting the geometry of the \ac{CMS} detector. The model discriminates between signal events of R-parity violating super-symmetry and background events with multiple Jets resulting from inelastic \ac{QCD} scattering (QCD multi-Jets). With the \ac{CNN} model, they achieved 1.85 times higher efficiency and 1.2 times higher expected significance compared to the traditional cut-based method. They demonstrated the scalability of the model at a large scale using \ac{HPC} resources with up to 1024 nodes.
The authors in \cite{chakraborty2019interpretable} proposing an interpretable network for multi-Jet classification using the Jet spectrum, termed S2(R), derived from a Taylor series of an arbitrary Jet \ac{MLP} classifier function. The network's intermediate feature is an infrared and collinear safe variables, named C-correlator, estimating the importance of S2(R) deposits at angular scales. It offers comparable performance to \acp{CNN} with simpler architecture and fewer inputs.  The paper \cite{liang2024jet} proposes a jet origin identification method for the electron-positron Higgs factory, classifying Jets into 11 categories:  5 quark species, 5 anti-quarks, and gluons. It achieves jet tagging efficiencies ranging from 67\% to 92\% and charge flip rates between 7\% and 24\%, utilizing the ParticleNet model. The method benefits jet physics and \ac{HEP} by enhancing rare Higgs decay measurements. It reduces QCD backgrounds and improves flavor tagging, crucial for Higgs boson property studies. The dataset consists of simulated $\nu \bar{\nu}  H, H \rightarrow jj$ events at 240 GeV, generated with a Geant4-based detector simulation. The best reported performance includes a 92\% efficiency for $b$-jets and a 7\% charge flip rate for charm quarks.

\color{black}

\subsection{Jet tracking}

\begin{figure*}
    \centering
    \includegraphics{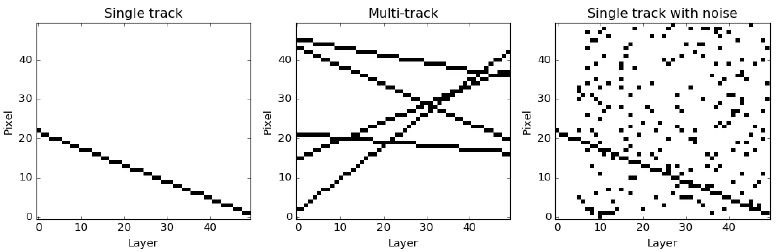}
    \caption{A toy dataset with adjustable dimensions, straight line representations for tracks, and the option to include uniform noise hits, all on a smaller scale.}
    \label{fig:track}
\end{figure*}

Jet tracking involves reconstructing the trajectories and properties of particles within Jets, formed when quarks and gluons fragment. Accurate tracking is vital for particle physics analyses, aiding in discoveries, \ac{SM} measurements, and searches for new phenomena. Advanced algorithms, including pattern recognition and \ac{ML}, are employed for precise tracking in modern detectors. In this research paper \cite{farrell2017hep}, the authors present early attempts at applying \ac{ML} techniques to address particle tracking challenges. This area remains largely unexplored, and they have just scratched the surface. Nonetheless, certain \ac{DL} methods show promise. \acp{LSTM} were found to be effective in solving the hit assignment problem in both 2D and 3D scenarios using a sequence of detector layer measurements, potentially offering an alternative to the combinatorial Kalman Filter. \acp{CNN} demonstrated the ability to construct representations of detector data from the ground up, aiding in hit assignment and parameter/uncertainty estimation. Through the combination of \ac{LSTM} and \ac{CNN}, the authors showcased a potentially powerful end-to-end model capable of identifying a variable number of tracks within detector images. Figure \ref{fig:track} displays sample 2D data generated with various types of tracks, including single-track, multi-track, and single-track with uniform noise.

\subsection{Jet generation}
In order to study new physics phenomena at the \ac{LHC}, it is necessary to simulate Monte Carlo events for both new physics signals and backgrounds. This simulation helps predict the experimental data expected from collider experiments. However, generating the large number of simulated events required for data analysis is time-consuming and computationally intensive using existing algorithms. Additionally, accurately simulating how energetic particles interact with detector materials can be a time-consuming process. In \cite{de2017learning}, researchers proposed using \acp{GAN} to build LAGAN framework, that is trained to generate authentic radiation distributions from simulated collisions involving high-energy particles. The authors found that the generated Jet images exhibited a wide range of pixel brightness levels and accurately reproduced low-dimensional physical observables such as reconstructed Jet mass and n-subJettiness. However, the study also acknowledges the limitations of this method and presents an empirical validation of the image quality. With further improvement, this approach could lead to faster simulation of \acp{HEP} events. Physicists at the \ac{LHC} use complex simulations to predict experimental outcomes. Generating vast amounts of simulated data is costly, but crucial for technique development. Challenges include accurately modeling detectors and particle interactions. In \cite{paganini2018accelerating}, researchers proposed a  \ac{GAN}-nased model for fast, accurate simulation of electromagnetic calorimeters. Despite ongoing precision challenges, this solution offers significant speed-ups, up to 100,000×, promising savings in computing resources and advancing physics research at the \ac{LHC} and beyond.

\subsection{Case studies in Jet tagging and classification}

To provide a deeper insight into the applications of \ac{ML} and \ac{DL} techniques in jet classification for \ac{HEP}, this section explores three critical case studies: top quark tagging, Higgs boson tagging, and photon jet classification.

\begin{itemize}[leftmargin=6pt] 
   \item \textbf{Top quark tagging:} This process is essential for distinguishing boosted top quarks from background events involving light quarks and gluons. Boosted top quarks often decay into a collimated spray of particles, which requires advanced tagging techniques to identify effectively. The ATLAS open data provides a comprehensive dataset for top quark tagging studies. Additionally, simulation tools like Delphes and MadGraph are frequently used to generate top quark events. Recent methods, including ParticleNet \cite{qu2020jet} and LorentzNet \cite{gong2022efficient}, have achieved significant improvements in classification accuracy by leveraging point-cloud representations of jets. These models employ graph-based architectures and permutation-invariant structures to enhance the discrimination power. Metrics such as classification accuracy and \ac{AUC} have demonstrated significant improvements for top quark tagging using LorentzNet, achieving values exceeding 94\% and 98.6\%, respectively.

\item \textbf{Higgs boson tagging:} Is crucial for validating the \ac{SM} and investigating potential new physics phenomena. Higgs bosons decaying into b-quarks generate jet structures with distinctive substructure features, making them a key focus for tagging studies. Datasets such as the \ac{CMS} open data and the Higgs dataset from the University of California, Irvine \ac{ML} repository serve as valuable resources for developing tagging algorithms. Traditional methods like \acp{BDT} and modern approaches such as \acp{CNN} have been employed extensively. Furthermore, advanced architectures like \ac{LGN} \cite{bogatskiy2020lorentz} and ParticleNet \cite{qu2020jet} have demonstrated superior classification capabilities. By utilizing high-level kinematic features and \ac{DL} techniques, classification accuracies exceeding 92\% and \ac{AUC} values surpassing 96\% have been achieved, along with notable background suppression.

\item \textbf{Photon Jet classification:} Photon jet classification is a critical task for studying the quark-gluon plasma and distinguishing between direct photons and those originating from fragmentation processes. Quark-gluon datasets generated using PYTHIA8 simulations form the basis for training and evaluating classification models, with additional opportunities provided by \ac{CMS} open data for analyzing real collision events. Advanced models such as  \ac{EGNN} \cite{satorras2021n} and \ac{PCT} \cite{mikuni2021point} have demonstrated effectiveness in capturing the energy deposits and angular distributions of particles within jets. Notably, state-of-the-art methods, such as \ac{EGNN}, have demonstrated exceptional performance in photon jet classification tasks, achieving accuracies above 92\% and \ac{AUC} values exceeding 97\%.
\end{itemize}

\color{black}

\section{Future direction and outlook}
\label{sec7}
The future of \ac{ML} and \ac{DL} in \ac{HEP}, particularly in Jet analysis, is poised for transformative advancements. As researchers delve deeper into the petabyte-scale datasets generated by experiments like those at the \ac{LHC} and \ac{QCD}, the role of \ac{DL} becomes increasingly vital. The potential implications of \ac{QML}-baset Jet research for future particle physics experiments are significant. By demonstrating the effectiveness of \ac{QML} for Jet classification in section \ref{sec5.1}, this opens up new possibilities for improving the performance of particle physics experiments. Researchers could apply the suggested \ac{QML}-based approaches to Jet images and \acp{PC} to other \ac{HEP} problems, such as signal versus background separation, anomaly detection, and particle track reconstruction. Furthermore, \ac{QML}-based research on Jet tagging could pave the way for the development of new quantum algorithms and hardware that could be used to solve complex problems in particle physics and other fields. 

There are multiple other compelling aspects and potential extensions that warrant further exploration, which are outlined here. For examples, for \textbf{\textit{event-level analysis}},  a Jet, in essence, cannot be entirely separated from an event's remaining parts, yet "pure" Jets can be achieved through grooming techniques. The utility of color connections is notable in various scenarios. The exploration into how to effectively demonstrate these effects is important, as there is potential in enhancing event-level analysis. The \ac{RNN} approach, particularly \ac{RecNN}, is easily adaptable for event-level analysis due to its natural fit into larger hierarchical structures. Previous studies have examined event analysis focusing solely on Jets, utilizing simple \ac{RNN} chains to reconstruct events from Jets. When considering event-level implementation, structuring the entire event poses a significant challenge. Viewing each event as a structured data tree, where the entire event's information is encapsulated in the nodes' properties and their interconnections, is vital. Therefore, accurately representing each element and its connections within the event is crucial for developing neural network architectures. For \textbf{\textit{Jet  unsupervised learning}}, within the \ac{DNN} framework, adjusting Jet clustering could potentially enhance performance. Treating Jet finding as a minimization problem presents an intriguing perspective, making it appealing to incorporate Jet finding processes directly into event-level analysis. Another example, for \textit{\textbf{new physics phenomena}} often display distinctive patterns related to their particle spectrum and decay modes. For instance, supersymmetry (SUSY) events typically generate a high number of final states, presenting a more complex hierarchical structure, and may include several soft leptons in electroweakino searches. Investigating whether \acp{DNN} can more effectively accommodate such topologies is also a worthwhile endeavor \cite{cheng2018recursive}. Moreover, distinguishing between quark-initiated and gluon-initiated Jets is crucial in collider experiments like the \ac{LHC}. Discriminating between these Jets is challenging due to complex correlations in radiation patterns and non-perturbative effects like hadronization. \ac{AI} methods, such as deep generative models, offer promising solutions to address this challenge  \cite{komiske2017deep}. Moving forward, there is a notable scarcity of published research on the application of \ac{AE} for Jet image processing, highlighting an opportunity for researchers to explore this field further. The potential for \ac{AE} to significantly improve the separation of Jet images and \ac{PC} from background noise presents a promising area of study. By focusing on this niche, researchers can contribute to advancing our understanding and methodologies in particle physics, potentially leading to more accurate and efficient analysis techniques. 

The complexity and volume of the data necessitate sophisticated analytical techniques that \ac{DL} models, especially those based on \acp{CNN} and \acp{GNN}, are well-equipped to handle. These models excel in identifying intricate patterns and correlations within the data, making them invaluable for tasks such as Jet tagging, particle tracking, and event classification. Furthermore, the scalability of \ac{DL} models needs to be addressed to handle the increasing data rates from next-generation detectors and accelerators. Efficient training algorithms and model compression techniques will be essential for deploying these models in real-time analysis frameworks, enabling faster decision-making processes for data acquisition and retention. The future of \ac{DL} in \textbf{\textit{Jet energy progression and estimation}} promises enhanced precision and efficiency. Innovations will likely focus on developing more sophisticated neural network models that can accurately predict Jet energies in complex environments. Emphasis on real-time data analysis capabilities and integration with experimental workflows will be crucial, driving advancements in detecting and interpreting high-energy particle collisions more effectively and swiftly. The future of DL-based Jet\textbf{\textit{ anomaly detection}} in HEP lies in advancing unsupervised learning techniques to uncover new physics signals hidden in complex data. Innovations in model interpretability and real-time processing will enhance detection capabilities. Cross-disciplinary collaboration will drive these advancements, leading to breakthroughs in identifying rare phenomena and expanding our understanding of the fundamental constituents of the universe. Others applications such as \textbf{\textit{flavor tagging}}, \textbf{\textit{pileup mitigation}}, and the \textbf{\textit{reconstruction of decay chains}}. These DL-based Jet classification can help in distinguishing between different types of particles based on their energy deposition patterns, aiding in the precise determination of particle origins and decay pathways. Additionally, they can be used for enhancing signal-to-noise ratios in complex collision environments, improving the accuracy of particle trajectory tracking, and in the analysis of Jet substructure to identify specific decay processes, contributing to a deeper understanding of the underlying physics in high-energy collisions. The application of DL-based \ac{HEP} Jet for \textbf{\textit{tomography}} is promising. This approach has the potential to revolutionize how we visualize and analyze subatomic particles, offering unprecedented precision and insight. By leveraging DL techniques, researchers can improve the accuracy of tomographic reconstructions, enhancing our understanding of particle interactions and the fundamental structure of matter.

\Ac{TL}, encompassing all its forms, including techniques like fine-tuning and domain adaptation \cite{kheddar2023deepASR,himeur2023video}, is poised to revolutionize Jet \ac{HEP} applications by leveraging pre-trained models from vast datasets to enhance performance on specific tasks with limited data. This approach can significantly reduce computational costs and training times, making it ideal for adapting models to new experiments or rare phenomena. As \ac{HEP} experiments generate increasingly complex data, the ability to apply knowledge from one context to another will be invaluable for improving event classification, anomaly detection, and signal processing. Looking ahead, \ac{TL} will be crucial for efficiently extracting insights from new particle interactions and advancing our understanding of fundamental physics. Exploring advanced architectures as sources of prior knowledge, such as EfficientNet, \ac{ViT}, Swin Transformers, ConvNeXt, \acp{GNN}, neural ordinary differential equations (NODEs), physics-informed neural networks (PINNs), and and AutoML for architecture optimization, could offer substantial improvements to target models conducting AI-based Jet tasks \cite{macaluso2018pulling}. These \ac{SOTA} methods are better suited to handling the complexities of particle physics data compared to older architectures like AlexNet or VGG. Generalizing the top tagger to classify other boosted objects, such as W/Z bosons, Higgs bosons, and other particles, remains straightforward, and extending it to partially-merged and fully resolved tops could enhance background rejection.

Systematic errors are a significant concern in \ac{HEP} experiments, particularly in image classification tasks involving jet analysis. These errors can arise from various sources, including detector calibration inaccuracies, biases in data reconstruction, and environmental factors during data acquisition. Addressing these uncertainties is crucial for the reliability and accuracy of \ac{ML} models applied in \ac{HEP}. One approach to mitigating systematic errors is through systematics-aware learning, which involves developing models that account for potential biases in the data. For instance, Estrade et al. \cite{estrade2019systematic} discussed the importance of creating benchmarks that capture realistic cases of systematic errors in \ac{HEP} analysis to facilitate experimental comparisons of different techniques. Another strategy involves adversarial learning to eliminate systematic errors \cite{clavijo2021adversarial}. This paper discusses the application of adversarial domain adaptation in an unsupervised setting to reduce sample bias in supervised \ac{HEP} event classifier training. The authors utilize a neural network with a gradient reversal layer to simultaneously enable signal versus background event classification while minimizing differences in the network's response to background samples from different Monte Carlo models. Ghosh et al. \cite{PhysRevD.104.056026} proposed classifiers that are fully aware of uncertainties and their corresponding nuisance parameters, demonstrating that this approach can enhance sensitivity to parameters of interest. By incorporating uncertainty directly into the learning process, models can achieve better performance compared to traditional strategies that do not account for such uncertainties.

To further enhance the mitigation of systematic errors, future research should focus on integrating uncertainty quantification and robust optimization directly into the design of \ac{ML} architectures. This includes the development of hybrid models that combine traditional statistical techniques with modern \ac{ML} approaches to explicitly model and correct for systematic effects. Additionally, employing advanced simulation techniques that better mimic real-world data will help reduce discrepancies between training datasets and experimental observations. Efforts should also be directed toward leveraging transfer learning to adapt models trained on simulated data to real-world experimental conditions more effectively. Another promising avenue is the application of federated learning in \ac{HEP}, which enables collaborative training across multiple experimental datasets while preserving data privacy. This approach could be particularly effective in creating more generalized models that are less sensitive to dataset-specific biases. Finally, incorporating interpretability and explainability methods into systematic error analysis will help researchers better understand how models respond to uncertainties and biases, providing actionable insights to refine both experiments and \ac{ML} methodologies. Such advancements will ultimately ensure that \ac{ML} models in \ac{HEP} are robust, transparent, and ready for real-world applications.

\color{black}
\Ac{RL}, with all its variants \cite{gueriani2023deep,kheddar2024reinforcement}, in \ac{HEP}  Jet applications is set to open novel pathways for optimizing experimental setups and data analysis strategies. By leveraging \ac{RL}'s ability to learn optimal policies through interaction with an environment, future \ac{HEP}  experiments could see enhanced automation in event selection, detector alignment, and real-time data processing. The adaptability of \ac{RL} models to dynamic systems makes them particularly suited for managing the complexities of particle collision events. As the technology matures, integrating \ac{RL} into \ac{HEP}  could lead to significant advancements in experiment efficiency, discovery potential, and the ability to navigate vast datasets to uncover new physics phenomena. Additionally, \ac{FL}-based computer vision  \cite{himeur2023federated} presents a promising frontier for Jet images  applications, offering a pathway to harness collaborative model training while preserving data privacy and security. By distributing the learning process across multiple nodes, each holding its own subset of data, \ac{FL} enables a collective improvement of models without direct data sharing. This approach is particularly suited for \ac{HEP}  collaborations spread across global institutions, where data locality and privacy concerns can limit traditional centralized training methods. Advancements in \ac{FL} could lead to more robust, accurate models, enhancing our understanding of complex particle physics phenomena through cooperative, privacy-preserving analysis between different \acp{LHC}. 

The integration of \acp{LLM} and generative \ac{AI} \cite{kheddar2024transformers} into \ac{HEP} has the potential to enhance the precision of particle detection and characterization. By leveraging these advanced \ac{AI} models, researchers can identify subtle patterns and anomalies in Jet tagging that might be missed by conventional methods. This improved accuracy is crucial for discovering new particles or interactions that could lead to breakthroughs in our understanding of the universe. For example, in the search for dark matter or other exotic particles, detecting faint signals amid a noisy background is a significant challenge. Generative \ac{AI} can help by producing simulations that highlight these weak signals, allowing physicists to fine-tune their detection algorithms. Similarly, \acp{LLM} can assist by providing context and insight into these findings, suggesting potential theoretical implications and further areas of exploration. The application of \acp{LLM} and generative \ac{AI} in \ac{HEP} also promotes a more collaborative and interdisciplinary approach to research. By integrating \ac{AI} experts with physicists, new methodologies and tools can be developed that leverage the strengths of both fields. This collaboration can lead to the creation of more sophisticated models that are specifically tailored to the needs of \ac{HEP}. Furthermore, the insights gained from \ac{HEP} research using \ac{AI} can be applied to other fields, such as astrophysics, medical imaging, and materials science. This cross-pollination of ideas and techniques can drive innovation across multiple disciplines, leading to advancements that benefit a wide range of scientific endeavors.

\section{Conclusion}
\label{sec8}

Given the comprehensive assessment of \ac{ML} and \ac{DL} applications within the realm of \ac{HEP} presented in this survey, it is evident that these techniques have significantly impacted various aspects of \ac{HEP} experimentation and phenomenological studies. Through a detailed exploration of diverse \ac{DL} approaches, including their application to \ac{HEP}  classification, Jet particle analysis, and other pertinent areas, this paper has highlighted the potential of ML and \ac{DL} techniques to enhance our understanding of particle physics phenomena. The analysis undertaken throughout this survey underscores the importance of leveraging AI models tailored to \ac{HEP} images and \acp{PC}, as well as the significance of \ac{SOTA} ML and \ac{DL} techniques in advancing \ac{HEP} inquiries. Specifically, the review has elucidated the implications of these techniques for tasks such as Jet tagging, Jet tracking, and particle classification, shedding light on their capabilities and limitations in addressing key challenges within the field. As we reflect on the current status of \ac{HEP} grounded in \ac{DL} methodologies, it becomes evident that while significant progress has been made, there remain inherent challenges that must be addressed to fully harness the potential of these approaches. These challenges include issues related to data quality, model interpretability, and generalization to diverse experimental conditions. Nonetheless, the survey also identifies promising avenues for future research endeavors, such as the development of novel \ac{DL} architectures tailored to \ac{HEP} data and the integration of domain-specific knowledge to enhance the performance of learning models. By addressing the challenges and leveraging the opportunities highlighted in this survey, researchers can continue to push the boundaries of \ac{HEP} experimentation and pave the way for groundbreaking discoveries in particle physics using \ac{AI} techniques.

\printcredits

\section*{Declaration of competing interest}

The authors declare that they have no known competing financial interests or personal relationships that could have appeared to influence the work reported in this paper.

\section*{Data availability}
No data was used for the research described in the article.

\section*{Acknowledgement}
The first author acknowledges that the study was partially funded by the Algerian Ministry of Higher Education and Scientific Research (Grant No. PRFU-- A25N01UN2601 20230001).

\balance
\bibliographystyle{elsarticle-num}
\bibliography{references}

\end{document}